\documentclass[prl,reprint,superscriptaddress,showpacs,preprintnumbers,nobibnotes,amsmath,amssymb, aps,longbibliography]{revtex4-2}

\usepackage{cancel}
\usepackage{accents}
\usepackage{mciteplus,slashed}
\usepackage{amssymb,cancel,amsmath}
\usepackage{dcolumn}
\usepackage{bm}
\usepackage[caption=false]{subfig}
\usepackage{appendix}
\usepackage{feynmp-auto}
\usepackage[T1]{fontenc}	
\usepackage{csvsimple}
\usepackage[colorlinks=true,citecolor=blue,linkcolor=blue]{hyperref}
\usepackage[section]{placeins}
\usepackage[capitalise]{cleveref}
\usepackage{booktabs}
\usepackage{graphicx}
\usepackage{subfig}
\usepackage{mathrsfs}
\usepackage{multirow}
\usepackage[utf8]{inputenc}
\usepackage{siunitx}
\usepackage{comment}
\usepackage{gensymb}
\usepackage{enumitem}

\usepackage[dvipsnames]{xcolor}
\usepackage[normalem]{ulem}

\begin{document}

\title{Visible inelasticity as a probe of tau flavor content of astrophysical neutrinos}

\begin{abstract}

Astrophysical neutrinos provide a unique probe of neutrino flavor changes over cosmological baselines. 
While the tau component of the neutrino flux is expected to arise almost entirely from mixing, current measurements rely primarily on rare double-cascade signatures. 
We investigate a complementary method to measure the tau fraction using the visible inelasticity of starting track events in neutrino telescopes. 
Muonic decays of tau leptons produce tracks with systematically larger visible inelasticity than those from muon neutrino interactions, potentially enabling statistical separation of the two flavors. 
Using realistic IceCube exposures and detector performance, we show that this observable already yields competitive sensitivity to the tau-to-muon flux ratio, $R_{\tau\mu}$, achievable with existing data. 
This approach may further enable flavor measurements of individual sources and the selection of tau-enhanced source catalogs. 
Starting-track inelasticity thus provides a powerful and immediately accessible probe of astrophysical neutrino flavor and of potential physics beyond standard neutrino mixing.

\end{abstract}

\author{Alex~Y.~Wen}
\email{alexwen@fas.harvard.edu}
\affiliation{Department of Physics \& Laboratory for Particle Physics and Cosmology, Harvard University, Cambridge, MA 02138, USA}

\author{Carlos~A.~Arg{\"u}elles}
\email{carguelles@fas.harvard.edu}
\affiliation{Department of Physics \& Laboratory for Particle Physics and Cosmology, Harvard University, Cambridge, MA 02138, USA}

\author{Sergio~Palomares-Ruiz}
\email{sergiopr@ific.uv.es}
\affiliation{Instituto de Física Corpuscular (IFIC), CSIC-Universitat de València, Parc Científic UV, C/ Catedrático José Beltrán 2, E-46980 Paterna, Spain}

\maketitle

\section{Introduction}
\label{sec:intro}

Astrophysical neutrinos, produced beyond the Solar System, are powerful probes of both high-energy astrophysics and fundamental particle physics. 
They are expected to arise from standard processes such as pion decays in energetic environments (e.g., active galactic nuclei); yet the main sources of the observed flux remain uncertain. 
Nevertheless, their detection~\cite{IceCube:2013low} provides a unique opportunity: unlike other cosmic messengers, neutrinos carry flavor and undergo flavor changes via mixing. 
After propagating over cosmic distances, standard astrophysical sources and neutrino oscillations predict an approximately homogeneous flavor composition at Earth, $(\nu_e:\nu_\mu:\nu_\tau)\sim(1:1:1)$~\cite{Learned:1994wg}.
Therefore, flavor ratios are important observables for probing physics beyond standard predictions~\cite{Ahlers:2018mkf, Ackermann:2019cxh, Arguelles:2019rbn} and have been a primary focus of study since the initial detection of high-energy astrophysical neutrinos~\cite{Mena:2014sja, Watanabe:2014qua, Palomares-Ruiz:2015mka, Palladino:2015zua, IceCube:2015rro, IceCube:2015gsk}.

Among the known neutrino flavors, the tau component, $\nu_\tau$, is particularly distinctive~\cite{MammenAbraham:2022xoc}. 
Unlike $\nu_e$ and $\nu_\mu$, it is not expected to be produced in appreciable amounts at astrophysical sources, as standard cosmic mechanisms yield negligible primary $\nu_\tau$ fluxes~\cite{Athar:2000rx}. 
Consequently, the observation of a tau component in a homogeneous flavor composition constitutes a direct test of oscillation physics, since astrophysical $\nu_\tau$ are generated almost entirely through flavor mixing during propagation. 
In the literature, the tau fraction has often been reported as the tau-to-muon flux ratio, $R_{\tau\mu} \equiv \left(\phi_{\nu_\tau} + \phi_{\bar\nu_\tau}\right) / \left(\phi_{\nu_\mu} + \phi_{\bar\nu_\mu} \right)$, and we will adopt the same observable in this work.
Experimentally, the allowed variation in $R_{\tau\mu}$ arising from current uncertainties in the neutrino mixing matrix elements~\cite{deSalas:2020pgw, Esteban:2024eli, Capozzi:2025wyn} is smaller than that of its electron counterpart $R_{e\mu}$~\cite{Lipari:2007su, Bustamante:2015waa}, making $R_{\tau\mu}$ a desirable observable. Therefore, measurements of the tau flavor over cosmic baselines provide a direct probe of neutrino mixing and, more broadly, of neutrino propagation on the largest accessible length scales in the Universe.

\begin{figure}[t]%
    \centering
    \includegraphics[width=0.99\linewidth]{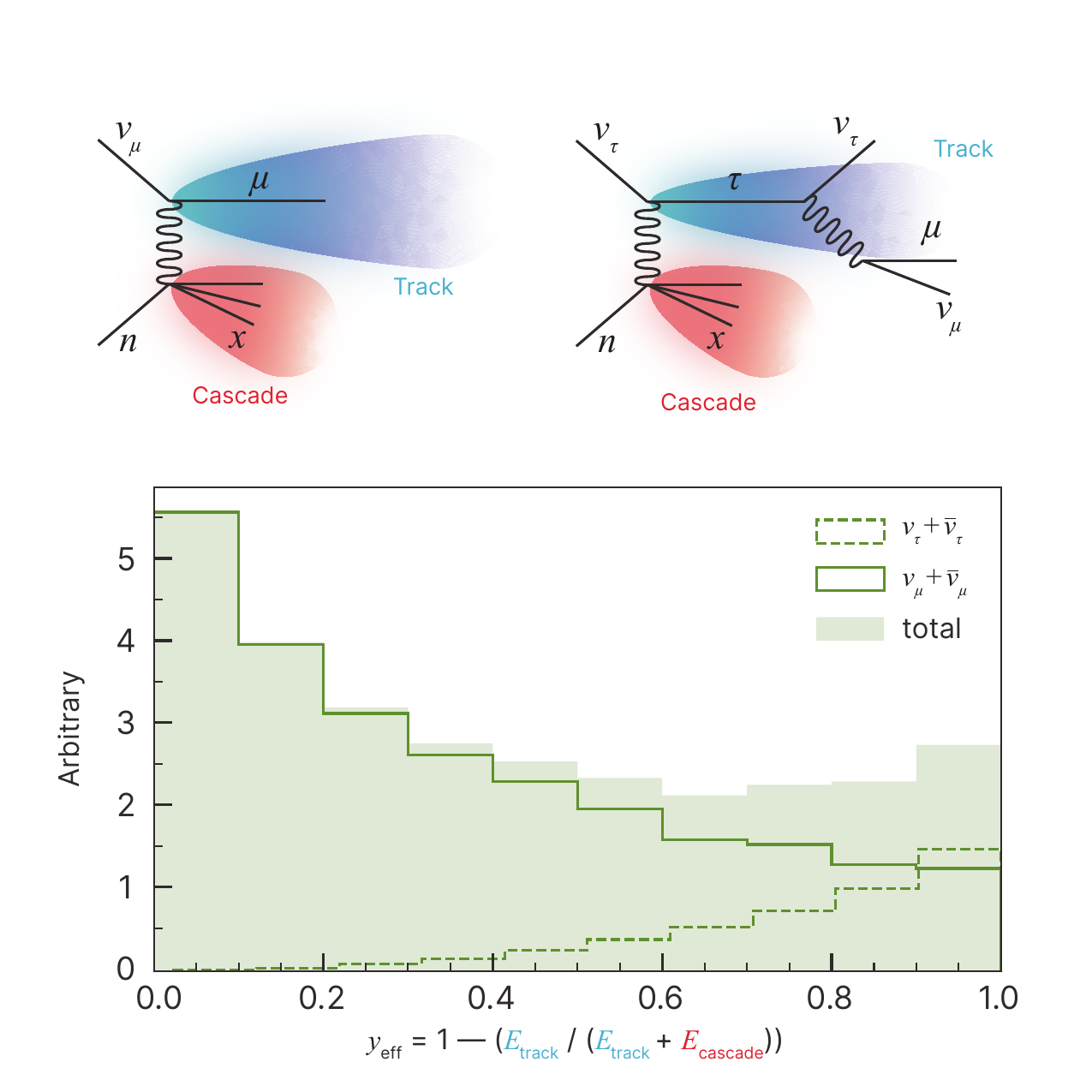}
    \caption{\textit{Flavored track signals.} Shown in the top of the figure are cartoons of $\nu_\mu$ (left) and $\nu_\tau$ (right) charged current interactions which result in a long, visible, track. In the $\nu_\mu$ case, the muon comes from the $\nu_\mu$ interaction, while in the $\nu_\tau$ case, the track comes from the tau and the subsequent muonic tau decay. In the bottom panel, we show the visible inelasticity, $y_\textrm{vis}$, distributions for the two types of tracks and their sum, for a representative flux of equal-flavor neutrinos distributed between $10^4\:\mathrm{GeV}$ and $10^6\:\mathrm{GeV}$ as $E_\nu^{-2.5}$.}%
    \label{fig:inelasticity_cartoon}%
\end{figure}

Significant departures from this tau-flavored expectation may signal new physics affecting flavor evolution beyond the established three-flavor mixing~\cite{Bhattacharya:2010xj, Arguelles:2015dca, Bustamante:2015waa, Shoemaker:2015qul, Rasmussen:2017ert, Song:2020nfh, Arguelles:2022tki}. 
For instance, neutrino decay~\cite{Beacom:2002vi, Maltoni:2008jr, Baerwald:2012kc, Bustamante:2016ciw, Abdullahi:2020rge}, quasi-Dirac neutrinos~\cite{Crocker:2001zs, Beacom:2003eu, Esmaili:2009fk, Carloni:2022cqz}, non-standard interactions~\cite{Blennow:2009rp, Gonzalez-Garcia:2016gpq}, coupling of neutrinos to dark matter~\cite{deSalas:2016svi, Farzan:2018pnk, Reynoso:2022vrn, Arguelles:2023wvf} or dark energy~\cite{Hung:2003jb, Ando:2009ts, Klop:2017dim}, sterile neutrinos~\cite{Athar:2000yw, Keranen:2003xd, Arguelles:2019tum}, Lorentz or CPT invariance violation~\cite{Barenboim:2003jm, Hooper:2005jp, Bhattacharya:2009tx, Bustamante:2010nq, Katori:2016eni}, violation of the equivalence principle~\cite{Minakata:1996nd}, and quantum decoherence~\cite{Hooper:2004xr, Anchordoqui:2005gj}, are all scenarios which may modify the flavor ratio to varying degrees.

For all the reasons above, the observation of tau neutrinos --- and the precise measurement of their content --- has long been an important experimental goal. 
Several techniques have been proposed to identify high-energy tau neutrino events in neutrino telescopes: \textit{double-bang} events~\cite{Learned1981DUMANDTau, Learned:1994wg} and \textit{double-pulse} events~\cite{Cowen:2007ny} (collectively referred to as \textit{double-cascade} events, including partially contained cases), \textit{lollipop} and \textit{inverted lollipop} events~\cite{Beacom:2003nh}, muonic tau decays~\cite{Bugaev:2003sw, DeYoung:2006fg} (including \textit{sugar daddy} and \textit{tautsie pop} events~\cite{Cowen:2007ny}), echo signals~\cite{Li:2016kra}, and identification of tau tracks~\cite{Kistler:2016ask}.
High-energy tau neutrinos, most likely of astrophysical origin, have successfully been observed with \textit{double-cascade} signatures at the IceCube neutrino telescope~\cite{IceCube:2020fpi, IceCube:2024nhk}.
This relies on the identification of two cascades, one from the tau neutrino interaction, and one from the hadronic tau decay. 
While successful, this type of search is technically challenging, because these cascades may be closely spaced together in space and time, and separating them is often challenging --- at energies below a few PeV, the tau track is too short and thus the events add to the cascade event rate. 
As a result, only a handful of tau candidates have been detected using this method, offering significant but still imprecise experimental constraints on the tau fraction. 

In this work, we focus on a complementary method to probe the tau flavor content of astrophysical neutrinos: using the visible inelasticity, $y_\mathrm{vis}$, as the primary discriminator. 
This necessarily requires a departure from the canonical double-cascade signature, and instead a focus on the \textit{starting-track} event signature. 
Indeed, this discriminator was already mentioned in the context of \textit{tautsie pop} events~\cite{Cowen:2007ny}, although it was not studied.
We will show that with current data at IceCube, precision on the scale of double-cascade measurements can already be achieved using this single-event morphology alone, highlighting the power of $y_\mathrm{vis}$ as a discriminating observable. 
Moreover, in addition to exploring a complementary way to measure $R_{\tau\mu}$, we discuss how the angular precision of tracks extends this method to point source flavor characterization and tau-enhanced source catalogs --- capabilities unavailable to cascade-based analyses.

\section{Inelasticity and Neutrino Telescopes}
\label{sec:inelasticity}

At energies from hundreds of GeV up to a few PeV, the primary experimental approach to detecting astrophysical neutrinos relies on ice- and water-based Cherenkov neutrino telescopes, such as IceCube~\cite{IceCube:2016zyt}, KM3NeT~\cite{KM3Net:2016zxf}, P-ONE~\cite{P-ONE:2020ljt}, Baikal-GVD~\cite{Avrorin:2015wba}, and TRIDENT~\cite{TRIDENT:2022hql}. 
At higher energies, alternative techniques based on Earth-skimming particle~\cite{TAMBO:2025jio}, optical~\cite{Otte:2025dld}, and radio~\cite{GRAND:2018iaj, GRAND:2025rps, RNO-G:2020rmc, PUEO:2020bnn, BEACON:2021fpe} detection become competitive. In this article we focus on the former ones, as they have already achieved neutrino detections.
Cherenkov neutrino telescopes consist of Gton-scale volumes of instrumented water or ice sensitive to the Cherenkov light emitted by charged particles produced in neutrino interactions, such as muons or hadronic shower components, that appear as signals in these detectors. 
Flavor information can be deduced from the unique morphologies of charged leptons produced in charged-current (CC) interactions: electrons create electromagnetic \textit{showers}, muons leave long \textit{tracks}, and taus typically decay hadronically but can also decay to a muon, leaving a track.
In particular, \textit{starting tracks} are produced by a lepton which was created by a neutrino CC interaction within the detector volume. 
This looks like a visible cascade at the interaction vertex, followed by a propagating lepton track.

The typical method for detecting tau neutrinos at neutrino telescopes has been to search for double cascades. 
As mentioned above, this is challenging because the tau decay length is typically very short, making the two cascades difficult to resolve, and consequently, the event rate has been low so far.
Additionally, for double bangs under 5~m, $D$-meson production in neutrino interactions is a significant background that limits tau identification~\cite{jin_2024_13235037}.

A complementary approach, explored in Refs.~\cite{Bugaev:2003sw, DeYoung:2006fg} for tau track identification, exploits the muonic tau decay, which occurs approximately $17.4\%$ of the time. 
When a tau neutrino interacts and the resulting tau decays to a muon, the event morphology resembles a muon-flavored starting track, but with a characteristically less energetic muon, since a significant fraction of the energy is carried away by invisible neutrinos in the tau decay. 
For a muon neutrino, by contrast, the outgoing muon carries a larger fraction of the neutrino energy, resulting in a less energetic cascade relative to the track.

The possible identification of the increase in brightness of the muon track from tau decay with respect to the tau track (even if the muon is less energetic) was considered in Refs.~\cite{Bugaev:2003sw, DeYoung:2006fg}, although reaching opposite conclusions. Here, to capture this distinction --- similarly to the suggestion in Ref.~\cite{Cowen:2007ny} for \textit{tautsie pop} events --- we study the \textit{visible inelasticity},
\begin{equation}
y_\mathrm{vis} = 1 - \frac{E_\mathrm{track}}{E_\mathrm{track}+E_\mathrm{cascade}} ~,
\end{equation}
which is closely related to the kinematic inelasticity $y$ used to describe deep inelastic scattering (DIS). 
Because the muon from a muonic tau decay is systematically less energetic than that from a muon neutrino interaction, the reconstructed $y_\mathrm{vis}$ is larger on average for tau-induced starting tracks. 
This difference in the $y_\mathrm{vis}$ distributions between the two flavors is illustrated in Fig.~\ref{fig:inelasticity_cartoon}, where the two distributions peak at distinctly different values. 
In Fig.~\ref{fig:inelasticity_cartoon}, a typical energy spectrum of $E_\nu^{-2.5}$ is assumed, but we have found the main conclusions to be generally robust to spectral index variations, especially in the allowed range suggested by modern astrophysical flux measurements~\cite{IceCube:2024fxo} (see Supplemental Material).
The visible inelasticity therefore provides a key statistical handle to discriminate tau and muon flavor components in a population of starting tracks and thereby, to measure $R_{\tau\mu}$. 
In the next section, we demonstrate, for the first time, the viability and sensitivity of such a measurement with a modern IceCube exposure.

\section{The Tau Fraction and Measurement Prospects}
\label{sec:taufraction}

Here, we assess the sensitivity of experiments like IceCube given an exposure approximately equivalent to 10 years of IceCube-sized detector data. 
To approximate this, we use the effective areas associated with a modern 10.3-year selection of starting-track events detailed in Ref.~\cite{IceCube:2024fxo}, from which we estimate the number of starting tracks available to perform an $R_{\tau\mu}$ measurement today.

Henceforth, in this work we assume an $E_\nu^{-2.58}$ energy spectrum, with the total rate of astrophysical events fixed by the all-charge, all-flavor, best-fit single power-law astrophysical flux reported in Ref.~\cite{IceCube:2024fxo},
\begin{equation} 
\label{eq:astroflux}
\begin{aligned} 
\phi_{\mathrm{astro}}(E_\nu) = 3 \times 1.68 & \times 10^{-18} \cdot  \left( E_\nu /100~\mathrm{TeV} \right)^{-2.58} \\ &
\mathrm{GeV}^{-1}\,\mathrm{cm}^{-2}\,\mathrm{s}^{-1}\,\mathrm{sr}^{-1} ~,
\end{aligned}
\end{equation}
in the energy range $(10 - 1000)~\mathrm{TeV}$. 
The lower bound avoids domination by the much softer atmospheric muon neutrino background. The upper bound avoids very long tau tracks. 
We also include the conventional atmospheric background, computed with the \texttt{MCEq} package~\cite{Fedynitch:2015zma} using the \texttt{SIBYLL2.3c} hadronic interaction model~\cite{Riehn:2017mfm} and the \texttt{HillasGaisser2012} cosmic-ray flux~\cite{Gaisser:2013bla}, which contributes significantly at low energies. We do not include, however, the subdominant contribution from atmospheric muons~\cite{IceCube:2018pgc}, which 
is expected to lie at the lowest-$y_\mathrm{vis}$ region.
Neutrinos are injected in an IceCube-sized volume using the \texttt{SIREN}~\cite{Schneider:2024eej} event generator according to the appropriate flux spectra. For all neutrinos arriving from the Southern sky, an approximate self-veto correction~\cite{Schonert:2008is, Gaisser:2014bja, Arguelles:2018awr} is implemented using the tables in Ref.~\cite{Arguelles:2018awr}, suppressing the atmospheric background.

\begin{figure}[t!]%
    \centering
    \includegraphics[width=0.99\linewidth]{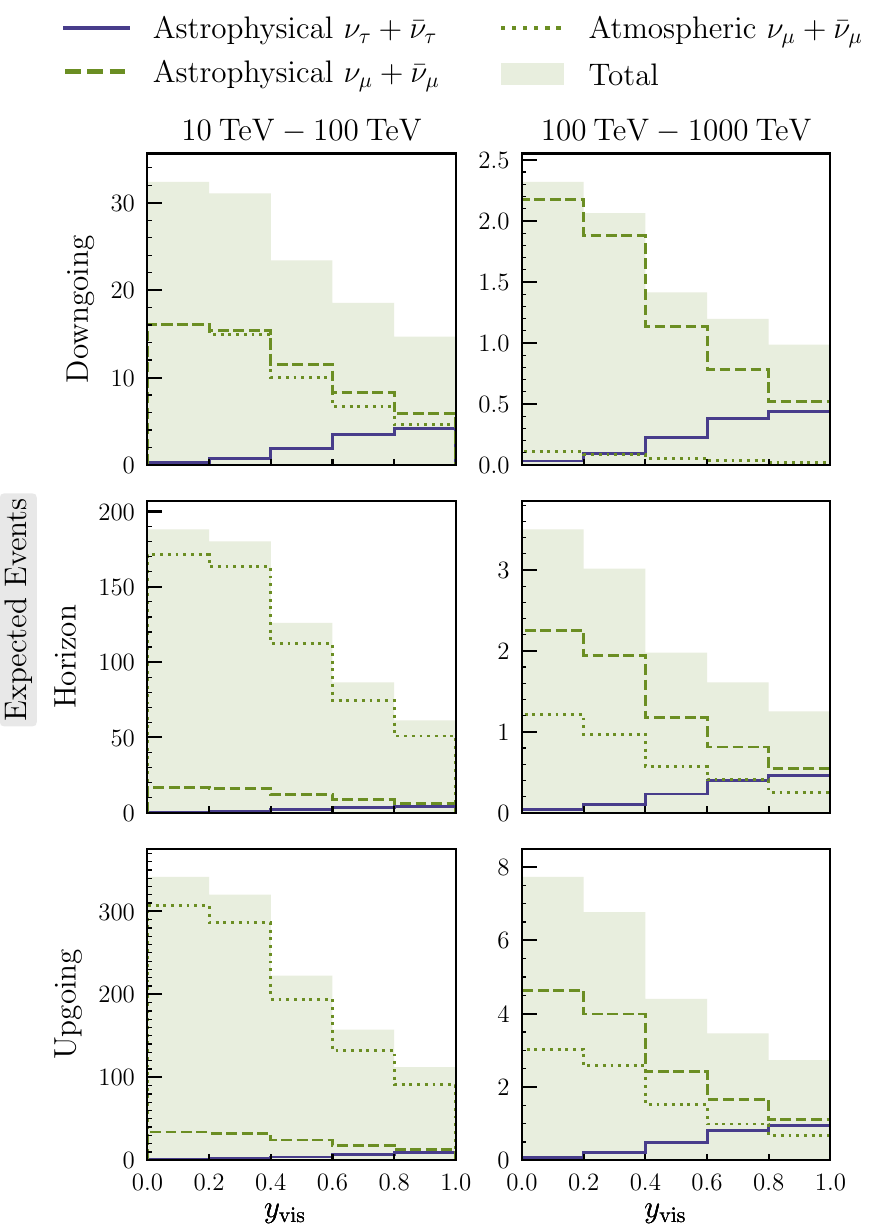}
    \caption{\textit{Event spectra as a function of the visible inelasticity, $y_\mathrm{vis}$}. We split into downgoing ($0.5 < \cos\theta_z < 1$, top row), horizon ($0 < \cos\theta_z < 0.5$, middle row), and upgoing ($-1 < \cos\theta_z < 0$, bottom row) zenith bins. We also split into two energy bins (the left and right columns) and the $y_\mathrm{vis}$ range ($0 < y_\mathrm{vis} < 1$) evenly into 5 bins. The $y_\mathrm{vis}$ value is smeared according to a Gaussian with standard deviation $\sigma_{y_\mathrm{vis}} = 0.2$. The astrophysical $\nu_\tau + \bar\nu_\tau$ component is more prominent at high $y_\mathrm{vis}$ and in the more downgoing regions, owing to the kinematics of the different-flavored decays and the self-veto suppression of the atmospheric neutrino background.} 
    \label{fig:event_distributions}%
\end{figure}

To evaluate the statistical sensitivity to $R_{\tau\mu}$, we partition events into: 1) two energy bins (10--100 TeV and 100--1000 TeV); 2) three zenith bins, $-1<\cos\theta_z<0$ (upgoing), $0<\cos\theta_z<0.5$ (horizontal), and $0.5<\cos\theta_z<1$ (downgoing); and 3) five $y_\mathrm{vis}$ bins equally spaced in $[0,1]$, yielding 30 bins in total. 
In each bin we assume Poisson-distributed contributions from atmospheric $\nu_\mu$, astrophysical $\nu_\mu$, and astrophysical $\nu_\tau$, with means given by integrating the corresponding fluxes, effective areas, and detector livetime. 
We use the true values of $E_\nu$ and $\cos\theta_z$, since for starting tracks the uncertainties on those quantities are significantly smaller than the bin size. 
For $y_\mathrm{vis}$, where the reconstruction uncertainty remains sizable, we apply a Gaussian smearing with a standard deviation of $\sigma_{ y_\mathrm{vis}} = 0.2$ across all energies. This corresponds to a realistic reconstruction performance that is possible with modern machine-learning-based methods~\cite{weigel_2025_15690410}.
In Fig.~\ref{fig:event_distributions} we show the binning configuration for the null scenario $R_{\tau\mu} = 1$ and the flux of Eq.~(\ref{eq:astroflux}) with the $y_\mathrm{vis}$ distribution shown in each subplot for a unique $(E_\nu, \cos\theta_z)$ bin. 
The astrophysical $\nu_\tau$ fraction varies significantly across all variables; the most discriminating power comes from bins at higher $\cos\theta_z$ (more downgoing) and higher $y_\mathrm{vis}$.

\begin{figure}[t!]%
    \centering
    \includegraphics[width=0.99\linewidth]{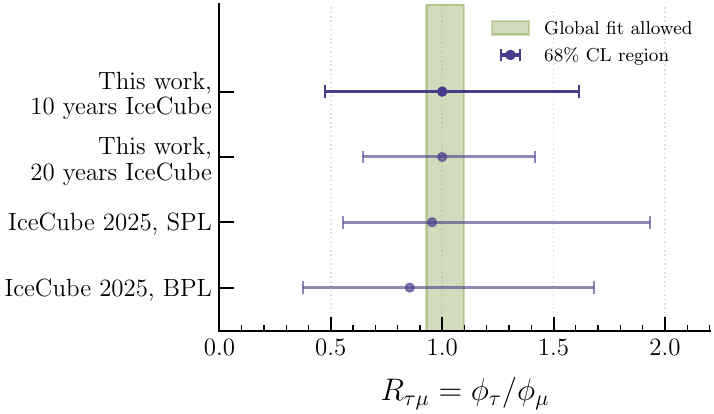}
    \caption{\textit{Projected all-sky flavor measurement sensitivity from visible inelasticity.} We show the 68\%~CL region of $R_{\tau\mu} \equiv \left(\phi_{\nu_\tau} + \phi_{\bar\nu_\tau}\right) / \left(\phi_{\nu_\mu} + \phi_{\bar\nu_\mu} \right)$ for our simulated all-sky sample of astrophysical neutrinos separated into zenith, inelasticity, and energy bins. Also shown for comparison is the profiled $R_{\tau\mu}$ 68\%~CL region from the IceCube MESE 2025 flavor measurement~\cite{IceCube:2025uyt}, for both single power law (SPL) and broken power law (BPL) astrophysical flux hypotheses. The vertical green band indicates the 68\% highest-density interval of $R_{\tau\mu}$ given NuFit2024 mixing parameter allowed ranges at the 68\%~CL~\cite{Esteban:2024eli}. Details on the binning are described in the main text and are shown in Fig.~\ref{fig:event_distributions}.}%
    \label{fig:sensitivity_diffuse}%
\end{figure}

We then evaluate the binned Poisson likelihood comparing an alternative model with varied $R_{\tau\mu}$ against the null hypothesis $R_{\tau\mu} = 1$. 
In addition, we profile over the overall normalization, $N_\phi$, and spectral index, $\gamma$.
The expected number of astrophysical events approximately depends on the overall normalization and spectral index through the combination $N_\phi \, \left(1 + 0.17 \, R_{\tau\mu}\right) \, \phi_{\nu_\mu}(\gamma)$, where the factor of $0.17$ accounts for the tau-to-muon decay braching ratio (see Supplemental Material).
We assume a Gaussian prior of 15\% on $N_\phi$, centered at 1. This represents an approximate precision level that can be achieved in astrophysical flux measurements~\cite{IceCube:2020wum, IceCube:2020acn, Abbasi:2021qfz}.
For completeness, the most conservative result assuming no prior on $N_\phi$ is shown in Suppl. Fig.~\ref{fig:sensitivity_diffuse_wtheory}.
We do not assume any prior on $\gamma$, since any reasonable prior does not have a significant impact on the sensitivity.
The resulting sensitivity is shown in Fig.~\ref{fig:sensitivity_diffuse} as the 68\% confidence level (CL) region on $R_{\tau\mu}$, obtained using Wilks' theorem with one degree of freedom. 
Also shown is the profiled 68\%~CL region from a recent lower-energy IceCube result that used all event morphologies, including double cascades, to infer the astrophysical flavor ratio~\cite{IceCube:2025uyt}. 
The competitiveness of our lower-statistics, single-morphology result highlights the discriminating power of $y_\mathrm{vis}$ in starting tracks. 
As a straightforward projection, we include the same measurement with twice the exposure, corresponding to about 20 years of IceCube data or a shorter period supplemented by another telescope such as Baikal-GVD or KM3NeT; the precise timeline for such a joint measurement will depend on the effective areas and reconstruction performance of those detectors, some of which are still under construction.

Finally, the solid green band in Fig.~\ref{fig:sensitivity_diffuse} shows the allowed region given uncertainties in both the neutrino mixing parameters and the source production hypotheses. 
We construct a posterior $P(\theta_{12},\theta_{23},\theta_{13},\delta_{CP},f_e)$ with priors on the mixing parameters taken from the NuFit 6.0 normal ordering likelihood maps~\cite{Esteban:2024eli}, including the joint $\theta_{23}$--$\delta_{CP}$ likelihood (which captures that respective correlation), and a uniform prior on $f_e$, which parameterizes the flavor ratio at sources as $(f_e, 1-f_e, 0)$ with the $f_e$ prior being uniform on $[0,1]$. 
The green band is the 68\% highest posterior density region of $P$, and can be easily compared to experimental projections and results.

\section{Discussion and Conclusions}
\label{sec:discussion}

While the experimental projections and bounds in Fig.~\ref{fig:sensitivity_diffuse} currently exceed the allowed region, as measurements improve, $R_{\tau\mu}$ could become one of the most sensitive probes of tau flavor, and any significant tension between measurement and theory would constitute major evidence for beyond Standard Model (BSM) effects in the neutrino sector.

A potential source of tau-flavored background are tau neutrinos from the prompt atmospheric flux. 
To assess their contribution, we compute the prompt flux using again the \texttt{SIBYLL2.3c} hadronic interaction model~\cite{Riehn:2017mfm} and the \texttt{HillasGaisser2012} cosmic ray flux~\cite{Gaisser:2013bla}. 
We find that a total of 1.6 prompt tau starting track events are expected across all zenith and energy ranges in this study in 10 years, with about 0.5 of those events expected in the downgoing direction ($\cos\theta_z > 0$). 
As such, the effects the prompt atmospheric neutrino background are expected to be negligibly small.

\begin{figure}[t!]%
    \centering
    \includegraphics[width=0.99\linewidth]{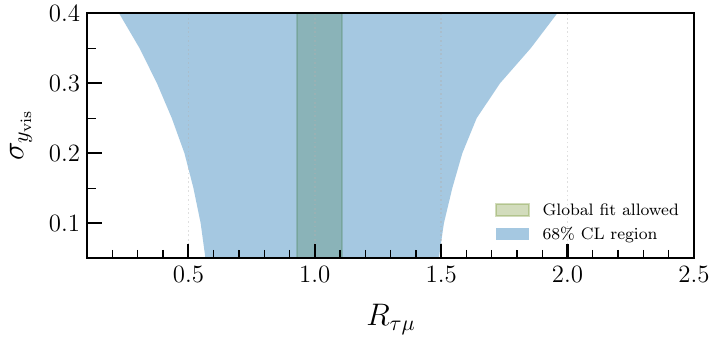}
    \caption{\textit{Effect of varying the $y_\textrm{vis}$ reconstruction quality.} The effect on $R_{\tau\mu}$ sensitivity, shown as the 68\%~CL region, of changing $\sigma_{y_\mathrm{vis}}$, the resolution on the visible inelasticity. 
    As a realistic benchmark, we consider $\sigma_{y_\mathrm{vis}} = 0.2$ as the nominal value used throughout this work.}%
    \label{fig:y-resolutions}%
\end{figure}

An important factor of this measurement is the ability to reconstruct the visible inelasticity $y_\mathrm{vis}$. 
Throughout this work we have assumed a resolution of $\sigma_{y_\mathrm{vis}} = 0.2$, a realistic value for modern machine-learning-based reconstructions in experiments such as IceCube~\cite{weigel_2025_15690410}. 
To study the sensitivity to this assumption, we have varied $\sigma_{y_\mathrm{vis}}$ between $0.05$ and $0.4$, and have evaluated the 68\%~CL region for $R_{\tau\mu}$ as in Fig.~\ref{fig:sensitivity_diffuse}, but increasing the $y_\mathrm{vis}$ binning to 10 equally-spaced bins to preserve shape information for smaller $\sigma_{y_\mathrm{vis}}$. 
The results are shown as the blue band in Fig.~\ref{fig:y-resolutions}. 
We find a moderate dependence on $y_\mathrm{vis}$ resolution: while the best achievable resolution should be pursued, even an excellent $\sigma_{y_\mathrm{vis}} = 0.05$ would not improve the experimental sensitivity to the level of the currently allowed region, so increased event statistics remain equally important.
In addition, a source of smearing of the $y_\mathrm{vis}$ distribution arises from final-state radiation, which is not included in this work. Yet, its effect on $y_\mathrm{vis}$ is expected to be at the percent level in the energy range we consider~\cite{Plestid:2024bva} and is therefore subdominant compared to the larger uncertainties associated with event reconstruction.

\begin{figure}[t!]%
    \centering
    \includegraphics[width=0.99\linewidth]{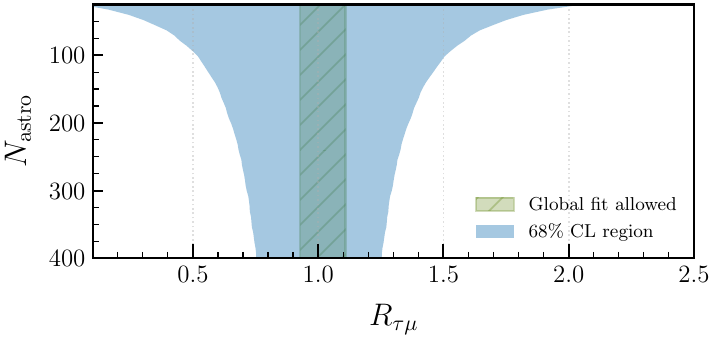}
    \caption{\textit{Sensitivity of the flavor-ratio measurement for a hypothetical point source.} The $R_{\tau\mu}$ sensitivity, shown is the 68\%~CL region, as a function of $N_\textrm{astro}$, the number of detected starting track events from a point source at $\cos\theta_z = 0.5$, assuming the same power-law flux as for the diffuse analysis. At these energies and assuming sub-degree angular uncertainty, the atmospheric background is largely irrelevant, resulting in a subleading dependence on the zenith angle.}%
    \label{fig:sensitivity_PS}%
\end{figure}

Using starting tracks for flavor measurements also enables point-source analyses, owing to the sub-degree angular resolution of track events — in contrast to the $\sim10^\circ$ resolution typical of cascades that dominate conventional flavor analyses in IceCube. 
Within the small angular window around a point source, the atmospheric background could become negligible, which would allow extending the analysis down to 1 TeV. 
In Fig.~\ref{fig:sensitivity_PS} we show the 68\%~CL region on $R_{\tau\mu}$ as a function of $N_\mathrm{astro}$, the number of detected events from a hypothetical source, binned as in Fig.~\ref{fig:event_distributions} but without the zenith angle dimension and with an additional energy bin, (1--10)~TeV. 
Even a statistically-limited flavor measurement of an individual source would be a major achievement: the flavor composition of individual astrophysical sources — aside from the Sun — remains entirely unexplored, and could directly connect flavor to known astrophysical processes. For next-generation detectors such as IceCube-Gen2~\cite{IceCube-Gen2:2020qha}, or from the combined exposure of all planned neutrino telescopes, such measurements may be within reach.

Finally, we highlight a further advantage of using tracks for flavor measurements: the construction of a tau-enhanced source catalog, again enabled by track angular precision. 
Current source searches at IceCube~\cite{IceCube:2022der} scan known catalogs of candidate sources — such as gamma-ray emitters — to minimize the trials factor of an all-sky search. 
A complementary approach is to build catalogs from tau-enriched events, which are predominantly astrophysical in origin. This principle already motivates dedicated detectors such as TAMBO~\cite{Thompson:2025tng}. 
With the starting-track measurement proposed in this work, a tau-enhanced sample could be constructed simply by retaining high-$y_\mathrm{vis}$ events: over a third of all downgoing events with $y_\mathrm{vis} > 0.8$ may be tau-flavored. 
A source search within this concentrated, high-purity catalog could be particularly fruitful for detectors with high sensitivity to Southern-sky events, such as KM3NeT, Baikal-GVD, P-ONE, or TRIDENT.

In this work we have shown that the muonic tau decay channel offers a viable and powerful path to flavor measurements at neutrino telescopes today, using already-collected samples of starting-track events. 
The observable $y_\mathrm{vis}$ provides a clean, statistically competitive probe of the astrophysical flavor ratio $R_{\tau\mu}$ that is sensitive to BSM physics in the neutrino sector. 
Looking ahead, the combination of growing exposures, improving $y_\mathrm{vis}$ reconstruction, and a new generation of neutrino telescopes will sharpen this measurement considerably. 
With the potential to resolve individual source flavors, construct tau-enriched catalogs, and test the standard neutrino mixing paradigm at astrophysical baselines, starting-track flavor measurements represent one of the most versatile and high-impact science cases for the coming era of multi-telescope neutrino astronomy.

\section*{Acknowledgements}

We thank Jesse Osborn for supplying the digitized starting track effective areas used in the analysis of Ref.~\cite{IceCube:2024fxo}, and Aswathi Balagopal V. for supplying the digitized flavor triangle result in Ref.~\cite{IceCube:2025uyt}.  
AYW is supported by the CIFAR SNSF Collaborative Research Catalyst Funds, National Science Foundation (NSF), and the Natural Sciences and Engineering Research Council of Canada (NSERC), funding reference number PGSD-577971-2023.
CAA are supported by the Faculty of Arts and Sciences of Harvard University, Canadian Institute for Advanced Research (CIFAR), the National Science Foundation (NSF), the John Templeton Foundation, the Research Corporation for Science Advancement, and the David \& Lucile Packard Foundation.
SPR is supported by grant PID2023-151418NB-I00, which is funded by MCIU/AEI/10.13039/501100011033/ FEDER, UE; by the Generalitat Valenciana grant CIPROM/2022/36; and by the European Union’s Horizon 2020 research and innovation programme under the Marie Skłodowska-Curie grant HORIZON-MSCA-2021-SE-01/101086085-ASYMMETRY.

\bibliography{main}

\begin{thebibliography}{98}%
\makeatletter
\providecommand \@ifxundefined [1]{%
 \@ifx{#1\undefined}
}%
\providecommand \@ifnum [1]{%
 \ifnum #1\expandafter \@firstoftwo
 \else \expandafter \@secondoftwo
 \fi
}%
\providecommand \@ifx [1]{%
 \ifx #1\expandafter \@firstoftwo
 \else \expandafter \@secondoftwo
 \fi
}%
\providecommand \natexlab [1]{#1}%
\providecommand \enquote  [1]{``#1''}%
\providecommand \bibnamefont  [1]{#1}%
\providecommand \bibfnamefont [1]{#1}%
\providecommand \citenamefont [1]{#1}%
\providecommand \href@noop [0]{\@secondoftwo}%
\providecommand \href [0]{\begingroup \@sanitize@url \@href}%
\providecommand \@href[1]{\@@startlink{#1}\@@href}%
\providecommand \@@href[1]{\endgroup#1\@@endlink}%
\providecommand \@sanitize@url [0]{\catcode `\\12\catcode `\$12\catcode `\&12\catcode `\#12\catcode `\^12\catcode `\_12\catcode `\%12\relax}%
\providecommand \@@startlink[1]{}%
\providecommand \@@endlink[0]{}%
\providecommand \url  [0]{\begingroup\@sanitize@url \@url }%
\providecommand \@url [1]{\endgroup\@href {#1}{\urlprefix }}%
\providecommand \urlprefix  [0]{URL }%
\providecommand \Eprint [0]{\href }%
\providecommand \doibase [0]{https://doi.org/}%
\providecommand \selectlanguage [0]{\@gobble}%
\providecommand \bibinfo  [0]{\@secondoftwo}%
\providecommand \bibfield  [0]{\@secondoftwo}%
\providecommand \translation [1]{[#1]}%
\providecommand \BibitemOpen [0]{}%
\providecommand \bibitemStop [0]{}%
\providecommand \bibitemNoStop [0]{.\EOS\space}%
\providecommand \EOS [0]{\spacefactor3000\relax}%
\providecommand \BibitemShut  [1]{\csname bibitem#1\endcsname}%
\let\auto@bib@innerbib\@empty
\bibitem [{\citenamefont {Aartsen}\ \emph {et~al.}(2013)\citenamefont {Aartsen} \emph {et~al.}}]{IceCube:2013low}%
  \BibitemOpen
  \bibfield  {author} {\bibinfo {author} {\bibfnamefont {M.~G.}\ \bibnamefont {Aartsen}} \emph {et~al.} (\bibinfo {collaboration} {IceCube Collaboration}),\ }\bibfield  {title} {\bibinfo {title} {{Evidence for high-energy extraterrestrial neutrinos at the IceCube detector}},\ }\href {https://doi.org/10.1126/science.1242856} {\bibfield  {journal} {\bibinfo  {journal} {Science}\ }\textbf {\bibinfo {volume} {342}},\ \bibinfo {pages} {1242856} (\bibinfo {year} {2013})},\ \Eprint {https://arxiv.org/abs/1311.5238} {arXiv:1311.5238 [astro-ph.HE]} \BibitemShut {NoStop}%
\bibitem [{\citenamefont {Learned}\ and\ \citenamefont {Pakvasa}(1995)}]{Learned:1994wg}%
  \BibitemOpen
  \bibfield  {author} {\bibinfo {author} {\bibfnamefont {J.~G.}\ \bibnamefont {Learned}}\ and\ \bibinfo {author} {\bibfnamefont {S.}~\bibnamefont {Pakvasa}},\ }\bibfield  {title} {\bibinfo {title} {{Detecting tau-neutrino oscillations at PeV energies}},\ }\href {https://doi.org/10.1016/0927-6505(94)00043-3} {\bibfield  {journal} {\bibinfo  {journal} {Astropart. Phys.}\ }\textbf {\bibinfo {volume} {3}},\ \bibinfo {pages} {267} (\bibinfo {year} {1995})},\ \Eprint {https://arxiv.org/abs/hep-ph/9405296} {arXiv:hep-ph/9405296} \BibitemShut {NoStop}%
\bibitem [{\citenamefont {Ahlers}\ \emph {et~al.}(2018)\citenamefont {Ahlers}, \citenamefont {Helbing},\ and\ \citenamefont {P{\'e}rez de~los Heros}}]{Ahlers:2018mkf}%
  \BibitemOpen
  \bibfield  {author} {\bibinfo {author} {\bibfnamefont {M.}~\bibnamefont {Ahlers}}, \bibinfo {author} {\bibfnamefont {K.}~\bibnamefont {Helbing}},\ and\ \bibinfo {author} {\bibfnamefont {C.}~\bibnamefont {P{\'e}rez de~los Heros}},\ }\bibfield  {title} {\bibinfo {title} {{Probing particle physics with IceCube}},\ }\href {https://doi.org/10.1140/epjc/s10052-018-6369-9} {\bibfield  {journal} {\bibinfo  {journal} {Eur. Phys. J. C}\ }\textbf {\bibinfo {volume} {78}},\ \bibinfo {pages} {924} (\bibinfo {year} {2018})},\ \Eprint {https://arxiv.org/abs/1806.05696} {arXiv:1806.05696 [astro-ph.HE]} \BibitemShut {NoStop}%
\bibitem [{\citenamefont {Ackermann}\ \emph {et~al.}(2019)\citenamefont {Ackermann} \emph {et~al.}}]{Ackermann:2019cxh}%
  \BibitemOpen
  \bibfield  {author} {\bibinfo {author} {\bibfnamefont {M.}~\bibnamefont {Ackermann}} \emph {et~al.},\ }\bibfield  {title} {\bibinfo {title} {{Fundamental physics with high-energy cosmic neutrinos}},\ }\href@noop {} {\bibfield  {journal} {\bibinfo  {journal} {Bull. Am. Astron. Soc.}\ }\textbf {\bibinfo {volume} {51}},\ \bibinfo {pages} {215} (\bibinfo {year} {2019})},\ \Eprint {https://arxiv.org/abs/1903.04333} {arXiv:1903.04333 [astro-ph.HE]} \BibitemShut {NoStop}%
\bibitem [{\citenamefont {Arg\"uelles}\ \emph {et~al.}(2020{\natexlab{a}})\citenamefont {Arg\"uelles}, \citenamefont {Bustamante}, \citenamefont {Kheirandish}, \citenamefont {Palomares-Ruiz}, \citenamefont {Salvado},\ and\ \citenamefont {Vincent}}]{Arguelles:2019rbn}%
  \BibitemOpen
  \bibfield  {author} {\bibinfo {author} {\bibfnamefont {C.~A.}\ \bibnamefont {Arg\"uelles}}, \bibinfo {author} {\bibfnamefont {M.}~\bibnamefont {Bustamante}}, \bibinfo {author} {\bibfnamefont {A.}~\bibnamefont {Kheirandish}}, \bibinfo {author} {\bibfnamefont {S.}~\bibnamefont {Palomares-Ruiz}}, \bibinfo {author} {\bibfnamefont {J.}~\bibnamefont {Salvado}},\ and\ \bibinfo {author} {\bibfnamefont {A.~C.}\ \bibnamefont {Vincent}},\ }\bibfield  {title} {\bibinfo {title} {{Fundamental physics with high-energy cosmic neutrinos today and in the future}},\ }\href {https://doi.org/10.22323/1.358.0849} {\bibfield  {journal} {\bibinfo  {journal} {PoS}\ }\textbf {\bibinfo {volume} {ICRC2019}},\ \bibinfo {pages} {849} (\bibinfo {year} {2020}{\natexlab{a}})},\ \Eprint {https://arxiv.org/abs/1907.08690} {arXiv:1907.08690 [astro-ph.HE]} \BibitemShut {NoStop}%
\bibitem [{\citenamefont {Mena}\ \emph {et~al.}(2014)\citenamefont {Mena}, \citenamefont {Palomares-Ruiz},\ and\ \citenamefont {Vincent}}]{Mena:2014sja}%
  \BibitemOpen
  \bibfield  {author} {\bibinfo {author} {\bibfnamefont {O.}~\bibnamefont {Mena}}, \bibinfo {author} {\bibfnamefont {S.}~\bibnamefont {Palomares-Ruiz}},\ and\ \bibinfo {author} {\bibfnamefont {A.~C.}\ \bibnamefont {Vincent}},\ }\bibfield  {title} {\bibinfo {title} {{Flavor composition of the high-energy neutrino events in IceCube}},\ }\href {https://doi.org/10.1103/PhysRevLett.113.091103} {\bibfield  {journal} {\bibinfo  {journal} {Phys. Rev. Lett.}\ }\textbf {\bibinfo {volume} {113}},\ \bibinfo {pages} {091103} (\bibinfo {year} {2014})},\ \Eprint {https://arxiv.org/abs/1404.0017} {arXiv:1404.0017 [astro-ph.HE]} \BibitemShut {NoStop}%
\bibitem [{\citenamefont {Watanabe}(2015)}]{Watanabe:2014qua}%
  \BibitemOpen
  \bibfield  {author} {\bibinfo {author} {\bibfnamefont {A.}~\bibnamefont {Watanabe}},\ }\bibfield  {title} {\bibinfo {title} {{The spectrum and flavor composition of the astrophysical neutrinos in IceCube}},\ }\href {https://doi.org/10.1088/1475-7516/2015/08/030} {\bibfield  {journal} {\bibinfo  {journal} {JCAP}\ }\textbf {\bibinfo {volume} {08}},\ \bibinfo {pages} {030}},\ \Eprint {https://arxiv.org/abs/1412.8264} {arXiv:1412.8264 [astro-ph.HE]} \BibitemShut {NoStop}%
\bibitem [{\citenamefont {Palomares-Ruiz}\ \emph {et~al.}(2015)\citenamefont {Palomares-Ruiz}, \citenamefont {Vincent},\ and\ \citenamefont {Mena}}]{Palomares-Ruiz:2015mka}%
  \BibitemOpen
  \bibfield  {author} {\bibinfo {author} {\bibfnamefont {S.}~\bibnamefont {Palomares-Ruiz}}, \bibinfo {author} {\bibfnamefont {A.~C.}\ \bibnamefont {Vincent}},\ and\ \bibinfo {author} {\bibfnamefont {O.}~\bibnamefont {Mena}},\ }\bibfield  {title} {\bibinfo {title} {{Spectral analysis of the high-energy IceCube neutrinos}},\ }\href {https://doi.org/10.1103/PhysRevD.91.103008} {\bibfield  {journal} {\bibinfo  {journal} {Phys. Rev. D}\ }\textbf {\bibinfo {volume} {91}},\ \bibinfo {pages} {103008} (\bibinfo {year} {2015})},\ \Eprint {https://arxiv.org/abs/1502.02649} {arXiv:1502.02649 [astro-ph.HE]} \BibitemShut {NoStop}%
\bibitem [{\citenamefont {Palladino}\ \emph {et~al.}(2015)\citenamefont {Palladino}, \citenamefont {Pagliaroli}, \citenamefont {Villante},\ and\ \citenamefont {Vissani}}]{Palladino:2015zua}%
  \BibitemOpen
  \bibfield  {author} {\bibinfo {author} {\bibfnamefont {A.}~\bibnamefont {Palladino}}, \bibinfo {author} {\bibfnamefont {G.}~\bibnamefont {Pagliaroli}}, \bibinfo {author} {\bibfnamefont {F.~L.}\ \bibnamefont {Villante}},\ and\ \bibinfo {author} {\bibfnamefont {F.}~\bibnamefont {Vissani}},\ }\bibfield  {title} {\bibinfo {title} {{What is the flavor of the cosmic neutrinos seen by IceCube?}},\ }\href {https://doi.org/10.1103/PhysRevLett.114.171101} {\bibfield  {journal} {\bibinfo  {journal} {Phys. Rev. Lett.}\ }\textbf {\bibinfo {volume} {114}},\ \bibinfo {pages} {171101} (\bibinfo {year} {2015})},\ \Eprint {https://arxiv.org/abs/1502.02923} {arXiv:1502.02923 [astro-ph.HE]} \BibitemShut {NoStop}%
\bibitem [{\citenamefont {Aartsen}\ \emph {et~al.}(2015{\natexlab{a}})\citenamefont {Aartsen} \emph {et~al.}}]{IceCube:2015rro}%
  \BibitemOpen
  \bibfield  {author} {\bibinfo {author} {\bibfnamefont {M.~G.}\ \bibnamefont {Aartsen}} \emph {et~al.} (\bibinfo {collaboration} {IceCube Collaboration}),\ }\bibfield  {title} {\bibinfo {title} {{Flavor ratio of astrophysical neutrinos above 35 TeV in IceCube}},\ }\href {https://doi.org/10.1103/PhysRevLett.114.171102} {\bibfield  {journal} {\bibinfo  {journal} {Phys. Rev. Lett.}\ }\textbf {\bibinfo {volume} {114}},\ \bibinfo {pages} {171102} (\bibinfo {year} {2015}{\natexlab{a}})},\ \Eprint {https://arxiv.org/abs/1502.03376} {arXiv:1502.03376 [astro-ph.HE]} \BibitemShut {NoStop}%
\bibitem [{\citenamefont {Aartsen}\ \emph {et~al.}(2015{\natexlab{b}})\citenamefont {Aartsen} \emph {et~al.}}]{IceCube:2015gsk}%
  \BibitemOpen
  \bibfield  {author} {\bibinfo {author} {\bibfnamefont {M.~G.}\ \bibnamefont {Aartsen}} \emph {et~al.} (\bibinfo {collaboration} {IceCube Collaboration}),\ }\bibfield  {title} {\bibinfo {title} {{A combined maximum-likelihood analysis of the high-energy astrophysical neutrino flux measured with IceCube}},\ }\href {https://doi.org/10.1088/0004-637X/809/1/98} {\bibfield  {journal} {\bibinfo  {journal} {Astrophys. J.}\ }\textbf {\bibinfo {volume} {809}},\ \bibinfo {pages} {98} (\bibinfo {year} {2015}{\natexlab{b}})},\ \Eprint {https://arxiv.org/abs/1507.03991} {arXiv:1507.03991 [astro-ph.HE]} \BibitemShut {NoStop}%
\bibitem [{\citenamefont {Mammen~Abraham}\ \emph {et~al.}(2022)\citenamefont {Mammen~Abraham} \emph {et~al.}}]{MammenAbraham:2022xoc}%
  \BibitemOpen
  \bibfield  {author} {\bibinfo {author} {\bibfnamefont {R.}~\bibnamefont {Mammen~Abraham}} \emph {et~al.},\ }\bibfield  {title} {\bibinfo {title} {{Tau neutrinos in the next decade: from GeV to EeV}},\ }\href {https://doi.org/10.1088/1361-6471/ac89d2} {\bibfield  {journal} {\bibinfo  {journal} {J. Phys. G}\ }\textbf {\bibinfo {volume} {49}},\ \bibinfo {pages} {110501} (\bibinfo {year} {2022})},\ \Eprint {https://arxiv.org/abs/2203.05591} {arXiv:2203.05591 [hep-ph]} \BibitemShut {NoStop}%
\bibitem [{\citenamefont {Athar}\ \emph {et~al.}(2000{\natexlab{a}})\citenamefont {Athar}, \citenamefont {Parente},\ and\ \citenamefont {Zas}}]{Athar:2000rx}%
  \BibitemOpen
  \bibfield  {author} {\bibinfo {author} {\bibfnamefont {H.}~\bibnamefont {Athar}}, \bibinfo {author} {\bibfnamefont {G.}~\bibnamefont {Parente}},\ and\ \bibinfo {author} {\bibfnamefont {E.}~\bibnamefont {Zas}},\ }\bibfield  {title} {\bibinfo {title} {{Prospects for observations of high-energy cosmic tau neutrinos}},\ }\href {https://doi.org/10.1103/PhysRevD.62.093010} {\bibfield  {journal} {\bibinfo  {journal} {Phys. Rev. D}\ }\textbf {\bibinfo {volume} {62}},\ \bibinfo {pages} {093010} (\bibinfo {year} {2000}{\natexlab{a}})},\ \Eprint {https://arxiv.org/abs/hep-ph/0006123} {arXiv:hep-ph/0006123} \BibitemShut {NoStop}%
\bibitem [{\citenamefont {de~Salas}\ \emph {et~al.}(2021)\citenamefont {de~Salas}, \citenamefont {Forero}, \citenamefont {Gariazzo}, \citenamefont {Mart{\'\i}nez-Mirav{\'e}}, \citenamefont {Mena}, \citenamefont {Ternes}, \citenamefont {T{\'o}rtola},\ and\ \citenamefont {Valle}}]{deSalas:2020pgw}%
  \BibitemOpen
  \bibfield  {author} {\bibinfo {author} {\bibfnamefont {P.~F.}\ \bibnamefont {de~Salas}}, \bibinfo {author} {\bibfnamefont {D.~V.}\ \bibnamefont {Forero}}, \bibinfo {author} {\bibfnamefont {S.}~\bibnamefont {Gariazzo}}, \bibinfo {author} {\bibfnamefont {P.}~\bibnamefont {Mart{\'\i}nez-Mirav{\'e}}}, \bibinfo {author} {\bibfnamefont {O.}~\bibnamefont {Mena}}, \bibinfo {author} {\bibfnamefont {C.~A.}\ \bibnamefont {Ternes}}, \bibinfo {author} {\bibfnamefont {M.}~\bibnamefont {T{\'o}rtola}},\ and\ \bibinfo {author} {\bibfnamefont {J.~W.~F.}\ \bibnamefont {Valle}},\ }\bibfield  {title} {\bibinfo {title} {{2020 global reassessment of the neutrino oscillation picture}},\ }\href {https://doi.org/10.1007/JHEP02(2021)071} {\bibfield  {journal} {\bibinfo  {journal} {JHEP}\ }\textbf {\bibinfo {volume} {02}},\ \bibinfo {pages} {071}},\ \Eprint {https://arxiv.org/abs/2006.11237} {arXiv:2006.11237 [hep-ph]} \BibitemShut {NoStop}%
\bibitem [{\citenamefont {Esteban}\ \emph {et~al.}(2024)\citenamefont {Esteban}, \citenamefont {Gonzalez-Garcia}, \citenamefont {Maltoni}, \citenamefont {Martinez-Soler}, \citenamefont {Pinheiro},\ and\ \citenamefont {Schwetz}}]{Esteban:2024eli}%
  \BibitemOpen
  \bibfield  {author} {\bibinfo {author} {\bibfnamefont {I.}~\bibnamefont {Esteban}}, \bibinfo {author} {\bibfnamefont {M.~C.}\ \bibnamefont {Gonzalez-Garcia}}, \bibinfo {author} {\bibfnamefont {M.}~\bibnamefont {Maltoni}}, \bibinfo {author} {\bibfnamefont {I.}~\bibnamefont {Martinez-Soler}}, \bibinfo {author} {\bibfnamefont {J.~P.}\ \bibnamefont {Pinheiro}},\ and\ \bibinfo {author} {\bibfnamefont {T.}~\bibnamefont {Schwetz}},\ }\bibfield  {title} {\bibinfo {title} {{NuFit-6.0: updated global analysis of three-flavor neutrino oscillations}},\ }\href {https://doi.org/10.1007/JHEP12(2024)216} {\bibfield  {journal} {\bibinfo  {journal} {JHEP}\ }\textbf {\bibinfo {volume} {12}},\ \bibinfo {pages} {216}},\ \Eprint {https://arxiv.org/abs/2410.05380} {arXiv:2410.05380 [hep-ph]} \BibitemShut {NoStop}%
\bibitem [{\citenamefont {Capozzi}\ \emph {et~al.}(2025)\citenamefont {Capozzi}, \citenamefont {Giar{\`e}}, \citenamefont {Lisi}, \citenamefont {Marrone}, \citenamefont {Melchiorri},\ and\ \citenamefont {Palazzo}}]{Capozzi:2025wyn}%
  \BibitemOpen
  \bibfield  {author} {\bibinfo {author} {\bibfnamefont {F.}~\bibnamefont {Capozzi}}, \bibinfo {author} {\bibfnamefont {W.}~\bibnamefont {Giar{\`e}}}, \bibinfo {author} {\bibfnamefont {E.}~\bibnamefont {Lisi}}, \bibinfo {author} {\bibfnamefont {A.}~\bibnamefont {Marrone}}, \bibinfo {author} {\bibfnamefont {A.}~\bibnamefont {Melchiorri}},\ and\ \bibinfo {author} {\bibfnamefont {A.}~\bibnamefont {Palazzo}},\ }\bibfield  {title} {\bibinfo {title} {{Neutrino masses and mixing: Entering the era of subpercent precision}},\ }\href {https://doi.org/10.1103/PhysRevD.111.093006} {\bibfield  {journal} {\bibinfo  {journal} {Phys. Rev. D}\ }\textbf {\bibinfo {volume} {111}},\ \bibinfo {pages} {093006} (\bibinfo {year} {2025})},\ \Eprint {https://arxiv.org/abs/2503.07752} {arXiv:2503.07752 [hep-ph]} \BibitemShut {NoStop}%
\bibitem [{\citenamefont {Lipari}\ \emph {et~al.}(2007)\citenamefont {Lipari}, \citenamefont {Lusignoli},\ and\ \citenamefont {Meloni}}]{Lipari:2007su}%
  \BibitemOpen
  \bibfield  {author} {\bibinfo {author} {\bibfnamefont {P.}~\bibnamefont {Lipari}}, \bibinfo {author} {\bibfnamefont {M.}~\bibnamefont {Lusignoli}},\ and\ \bibinfo {author} {\bibfnamefont {D.}~\bibnamefont {Meloni}},\ }\bibfield  {title} {\bibinfo {title} {{Flavor composition and energy spectrum of astrophysical neutrinos}},\ }\href {https://doi.org/10.1103/PhysRevD.75.123005} {\bibfield  {journal} {\bibinfo  {journal} {Phys. Rev. D}\ }\textbf {\bibinfo {volume} {75}},\ \bibinfo {pages} {123005} (\bibinfo {year} {2007})},\ \Eprint {https://arxiv.org/abs/0704.0718} {arXiv:0704.0718 [astro-ph]} \BibitemShut {NoStop}%
\bibitem [{\citenamefont {Bustamante}\ \emph {et~al.}(2015)\citenamefont {Bustamante}, \citenamefont {Beacom},\ and\ \citenamefont {Winter}}]{Bustamante:2015waa}%
  \BibitemOpen
  \bibfield  {author} {\bibinfo {author} {\bibfnamefont {M.}~\bibnamefont {Bustamante}}, \bibinfo {author} {\bibfnamefont {J.~F.}\ \bibnamefont {Beacom}},\ and\ \bibinfo {author} {\bibfnamefont {W.}~\bibnamefont {Winter}},\ }\bibfield  {title} {\bibinfo {title} {{Theoretically palatable flavor combinations of astrophysical neutrinos}},\ }\href {https://doi.org/10.1103/PhysRevLett.115.161302} {\bibfield  {journal} {\bibinfo  {journal} {Phys. Rev. Lett.}\ }\textbf {\bibinfo {volume} {115}},\ \bibinfo {pages} {161302} (\bibinfo {year} {2015})},\ \Eprint {https://arxiv.org/abs/1506.02645} {arXiv:1506.02645 [astro-ph.HE]} \BibitemShut {NoStop}%
\bibitem [{\citenamefont {Bhattacharya}\ \emph {et~al.}(2010{\natexlab{a}})\citenamefont {Bhattacharya}, \citenamefont {Choubey}, \citenamefont {Gandhi},\ and\ \citenamefont {Watanabe}}]{Bhattacharya:2010xj}%
  \BibitemOpen
  \bibfield  {author} {\bibinfo {author} {\bibfnamefont {A.}~\bibnamefont {Bhattacharya}}, \bibinfo {author} {\bibfnamefont {S.}~\bibnamefont {Choubey}}, \bibinfo {author} {\bibfnamefont {R.}~\bibnamefont {Gandhi}},\ and\ \bibinfo {author} {\bibfnamefont {A.}~\bibnamefont {Watanabe}},\ }\bibfield  {title} {\bibinfo {title} {{Ultra-high neutrino fluxes as a probe for non-standard physics}},\ }\href {https://doi.org/10.1088/1475-7516/2010/09/009} {\bibfield  {journal} {\bibinfo  {journal} {JCAP}\ }\textbf {\bibinfo {volume} {09}},\ \bibinfo {pages} {009}},\ \Eprint {https://arxiv.org/abs/1006.3082} {arXiv:1006.3082 [hep-ph]} \BibitemShut {NoStop}%
\bibitem [{\citenamefont {Arg{\"u}elles}\ \emph {et~al.}(2015)\citenamefont {Arg{\"u}elles}, \citenamefont {Katori},\ and\ \citenamefont {Salvado}}]{Arguelles:2015dca}%
  \BibitemOpen
  \bibfield  {author} {\bibinfo {author} {\bibfnamefont {C.~A.}\ \bibnamefont {Arg{\"u}elles}}, \bibinfo {author} {\bibfnamefont {T.}~\bibnamefont {Katori}},\ and\ \bibinfo {author} {\bibfnamefont {J.}~\bibnamefont {Salvado}},\ }\bibfield  {title} {\bibinfo {title} {{New physics in astrophysical neutrino flavor}},\ }\href {https://doi.org/10.1103/PhysRevLett.115.161303} {\bibfield  {journal} {\bibinfo  {journal} {Phys. Rev. Lett.}\ }\textbf {\bibinfo {volume} {115}},\ \bibinfo {pages} {161303} (\bibinfo {year} {2015})},\ \Eprint {https://arxiv.org/abs/1506.02043} {arXiv:1506.02043 [hep-ph]} \BibitemShut {NoStop}%
\bibitem [{\citenamefont {Shoemaker}\ and\ \citenamefont {Murase}(2016)}]{Shoemaker:2015qul}%
  \BibitemOpen
  \bibfield  {author} {\bibinfo {author} {\bibfnamefont {I.~M.}\ \bibnamefont {Shoemaker}}\ and\ \bibinfo {author} {\bibfnamefont {K.}~\bibnamefont {Murase}},\ }\bibfield  {title} {\bibinfo {title} {{Probing BSM neutrino physics with flavor and spectral distortions: prospects for future high-energy neutrino telescopes}},\ }\href {https://doi.org/10.1103/PhysRevD.93.085004} {\bibfield  {journal} {\bibinfo  {journal} {Phys. Rev. D}\ }\textbf {\bibinfo {volume} {93}},\ \bibinfo {pages} {085004} (\bibinfo {year} {2016})},\ \Eprint {https://arxiv.org/abs/1512.07228} {arXiv:1512.07228 [astro-ph.HE]} \BibitemShut {NoStop}%
\bibitem [{\citenamefont {Rasmussen}\ \emph {et~al.}(2017)\citenamefont {Rasmussen}, \citenamefont {Lechner}, \citenamefont {Ackermann}, \citenamefont {Kowalski},\ and\ \citenamefont {Winter}}]{Rasmussen:2017ert}%
  \BibitemOpen
  \bibfield  {author} {\bibinfo {author} {\bibfnamefont {R.~W.}\ \bibnamefont {Rasmussen}}, \bibinfo {author} {\bibfnamefont {L.}~\bibnamefont {Lechner}}, \bibinfo {author} {\bibfnamefont {M.}~\bibnamefont {Ackermann}}, \bibinfo {author} {\bibfnamefont {M.}~\bibnamefont {Kowalski}},\ and\ \bibinfo {author} {\bibfnamefont {W.}~\bibnamefont {Winter}},\ }\bibfield  {title} {\bibinfo {title} {{Astrophysical neutrinos flavored with beyond the Standard Model physics}},\ }\href {https://doi.org/10.1103/PhysRevD.96.083018} {\bibfield  {journal} {\bibinfo  {journal} {Phys. Rev. D}\ }\textbf {\bibinfo {volume} {96}},\ \bibinfo {pages} {083018} (\bibinfo {year} {2017})},\ \Eprint {https://arxiv.org/abs/1707.07684} {arXiv:1707.07684 [hep-ph]} \BibitemShut {NoStop}%
\bibitem [{\citenamefont {Song}\ \emph {et~al.}(2021)\citenamefont {Song}, \citenamefont {Li}, \citenamefont {Arg\"uelles}, \citenamefont {Bustamante},\ and\ \citenamefont {Vincent}}]{Song:2020nfh}%
  \BibitemOpen
  \bibfield  {author} {\bibinfo {author} {\bibfnamefont {N.}~\bibnamefont {Song}}, \bibinfo {author} {\bibfnamefont {S.~W.}\ \bibnamefont {Li}}, \bibinfo {author} {\bibfnamefont {C.~A.}\ \bibnamefont {Arg\"uelles}}, \bibinfo {author} {\bibfnamefont {M.}~\bibnamefont {Bustamante}},\ and\ \bibinfo {author} {\bibfnamefont {A.~C.}\ \bibnamefont {Vincent}},\ }\bibfield  {title} {\bibinfo {title} {{The future of high-energy astrophysical neutrino flavor measurements}},\ }\href {https://doi.org/10.1088/1475-7516/2021/04/054} {\bibfield  {journal} {\bibinfo  {journal} {JCAP}\ }\textbf {\bibinfo {volume} {04}},\ \bibinfo {pages} {054}},\ \Eprint {https://arxiv.org/abs/2012.12893} {arXiv:2012.12893 [hep-ph]} \BibitemShut {NoStop}%
\bibitem [{\citenamefont {Arg\"uelles}\ \emph {et~al.}(2023{\natexlab{a}})\citenamefont {Arg\"uelles} \emph {et~al.}}]{Arguelles:2022tki}%
  \BibitemOpen
  \bibfield  {author} {\bibinfo {author} {\bibfnamefont {C.~A.}\ \bibnamefont {Arg\"uelles}} \emph {et~al.},\ }\bibfield  {title} {\bibinfo {title} {{Snowmass white paper: beyond the standard model effects on neutrino flavor}},\ }\href {https://doi.org/10.1140/epjc/s10052-022-11049-7} {\bibfield  {journal} {\bibinfo  {journal} {Eur. Phys. J. C}\ }\textbf {\bibinfo {volume} {83}},\ \bibinfo {pages} {15} (\bibinfo {year} {2023}{\natexlab{a}})},\ \Eprint {https://arxiv.org/abs/2203.10811} {arXiv:2203.10811 [hep-ph]} \BibitemShut {NoStop}%
\bibitem [{\citenamefont {Beacom}\ \emph {et~al.}(2003{\natexlab{a}})\citenamefont {Beacom}, \citenamefont {Bell}, \citenamefont {Hooper}, \citenamefont {Pakvasa},\ and\ \citenamefont {Weiler}}]{Beacom:2002vi}%
  \BibitemOpen
  \bibfield  {author} {\bibinfo {author} {\bibfnamefont {J.~F.}\ \bibnamefont {Beacom}}, \bibinfo {author} {\bibfnamefont {N.~F.}\ \bibnamefont {Bell}}, \bibinfo {author} {\bibfnamefont {D.}~\bibnamefont {Hooper}}, \bibinfo {author} {\bibfnamefont {S.}~\bibnamefont {Pakvasa}},\ and\ \bibinfo {author} {\bibfnamefont {T.~J.}\ \bibnamefont {Weiler}},\ }\bibfield  {title} {\bibinfo {title} {{Decay of high-energy astrophysical neutrinos}},\ }\href {https://doi.org/10.1103/PhysRevLett.90.181301} {\bibfield  {journal} {\bibinfo  {journal} {Phys. Rev. Lett.}\ }\textbf {\bibinfo {volume} {90}},\ \bibinfo {pages} {181301} (\bibinfo {year} {2003}{\natexlab{a}})},\ \Eprint {https://arxiv.org/abs/hep-ph/0211305} {arXiv:hep-ph/0211305} \BibitemShut {NoStop}%
\bibitem [{\citenamefont {Maltoni}\ and\ \citenamefont {Winter}(2008)}]{Maltoni:2008jr}%
  \BibitemOpen
  \bibfield  {author} {\bibinfo {author} {\bibfnamefont {M.}~\bibnamefont {Maltoni}}\ and\ \bibinfo {author} {\bibfnamefont {W.}~\bibnamefont {Winter}},\ }\bibfield  {title} {\bibinfo {title} {{Testing neutrino oscillations plus decay with neutrino telescopes}},\ }\href {https://doi.org/10.1088/1126-6708/2008/07/064} {\bibfield  {journal} {\bibinfo  {journal} {JHEP}\ }\textbf {\bibinfo {volume} {07}},\ \bibinfo {pages} {064}},\ \Eprint {https://arxiv.org/abs/0803.2050} {arXiv:0803.2050 [hep-ph]} \BibitemShut {NoStop}%
\bibitem [{\citenamefont {Baerwald}\ \emph {et~al.}(2012)\citenamefont {Baerwald}, \citenamefont {Bustamante},\ and\ \citenamefont {Winter}}]{Baerwald:2012kc}%
  \BibitemOpen
  \bibfield  {author} {\bibinfo {author} {\bibfnamefont {P.}~\bibnamefont {Baerwald}}, \bibinfo {author} {\bibfnamefont {M.}~\bibnamefont {Bustamante}},\ and\ \bibinfo {author} {\bibfnamefont {W.}~\bibnamefont {Winter}},\ }\bibfield  {title} {\bibinfo {title} {{Neutrino decays over cosmological distances and the implications for neutrino telescopes}},\ }\href {https://doi.org/10.1088/1475-7516/2012/10/020} {\bibfield  {journal} {\bibinfo  {journal} {JCAP}\ }\textbf {\bibinfo {volume} {10}},\ \bibinfo {pages} {020}},\ \Eprint {https://arxiv.org/abs/1208.4600} {arXiv:1208.4600 [astro-ph.CO]} \BibitemShut {NoStop}%
\bibitem [{\citenamefont {Bustamante}\ \emph {et~al.}(2017)\citenamefont {Bustamante}, \citenamefont {Beacom},\ and\ \citenamefont {Murase}}]{Bustamante:2016ciw}%
  \BibitemOpen
  \bibfield  {author} {\bibinfo {author} {\bibfnamefont {M.}~\bibnamefont {Bustamante}}, \bibinfo {author} {\bibfnamefont {J.~F.}\ \bibnamefont {Beacom}},\ and\ \bibinfo {author} {\bibfnamefont {K.}~\bibnamefont {Murase}},\ }\bibfield  {title} {\bibinfo {title} {{Testing decay of astrophysical neutrinos with incomplete information}},\ }\href {https://doi.org/10.1103/PhysRevD.95.063013} {\bibfield  {journal} {\bibinfo  {journal} {Phys. Rev. D}\ }\textbf {\bibinfo {volume} {95}},\ \bibinfo {pages} {063013} (\bibinfo {year} {2017})},\ \Eprint {https://arxiv.org/abs/1610.02096} {arXiv:1610.02096 [astro-ph.HE]} \BibitemShut {NoStop}%
\bibitem [{\citenamefont {Abdullahi}\ and\ \citenamefont {Denton}(2020)}]{Abdullahi:2020rge}%
  \BibitemOpen
  \bibfield  {author} {\bibinfo {author} {\bibfnamefont {A.}~\bibnamefont {Abdullahi}}\ and\ \bibinfo {author} {\bibfnamefont {P.~B.}\ \bibnamefont {Denton}},\ }\bibfield  {title} {\bibinfo {title} {{Visible decay of astrophysical neutrinos at IceCube}},\ }\href {https://doi.org/10.1103/PhysRevD.102.023018} {\bibfield  {journal} {\bibinfo  {journal} {Phys. Rev. D}\ }\textbf {\bibinfo {volume} {102}},\ \bibinfo {pages} {023018} (\bibinfo {year} {2020})},\ \Eprint {https://arxiv.org/abs/2005.07200} {arXiv:2005.07200 [hep-ph]} \BibitemShut {NoStop}%
\bibitem [{\citenamefont {Crocker}\ \emph {et~al.}(2002)\citenamefont {Crocker}, \citenamefont {Melia},\ and\ \citenamefont {Volkas}}]{Crocker:2001zs}%
  \BibitemOpen
  \bibfield  {author} {\bibinfo {author} {\bibfnamefont {R.~M.}\ \bibnamefont {Crocker}}, \bibinfo {author} {\bibfnamefont {F.}~\bibnamefont {Melia}},\ and\ \bibinfo {author} {\bibfnamefont {R.~R.}\ \bibnamefont {Volkas}},\ }\bibfield  {title} {\bibinfo {title} {{Searching for long wavelength neutrino oscillations in the distorted neutrino spectrum of galactic supernova remnants}},\ }\href {https://doi.org/10.1086/340278} {\bibfield  {journal} {\bibinfo  {journal} {Astrophys. J. Suppl.}\ }\textbf {\bibinfo {volume} {141}},\ \bibinfo {pages} {147} (\bibinfo {year} {2002})},\ \Eprint {https://arxiv.org/abs/astro-ph/0106090} {arXiv:astro-ph/0106090} \BibitemShut {NoStop}%
\bibitem [{\citenamefont {Beacom}\ \emph {et~al.}(2004)\citenamefont {Beacom}, \citenamefont {Bell}, \citenamefont {Hooper}, \citenamefont {Learned}, \citenamefont {Pakvasa},\ and\ \citenamefont {Weiler}}]{Beacom:2003eu}%
  \BibitemOpen
  \bibfield  {author} {\bibinfo {author} {\bibfnamefont {J.~F.}\ \bibnamefont {Beacom}}, \bibinfo {author} {\bibfnamefont {N.~F.}\ \bibnamefont {Bell}}, \bibinfo {author} {\bibfnamefont {D.}~\bibnamefont {Hooper}}, \bibinfo {author} {\bibfnamefont {J.~G.}\ \bibnamefont {Learned}}, \bibinfo {author} {\bibfnamefont {S.}~\bibnamefont {Pakvasa}},\ and\ \bibinfo {author} {\bibfnamefont {T.~J.}\ \bibnamefont {Weiler}},\ }\bibfield  {title} {\bibinfo {title} {{PseudoDirac neutrinos: a challenge for neutrino telescopes}},\ }\href {https://doi.org/10.1103/PhysRevLett.92.011101} {\bibfield  {journal} {\bibinfo  {journal} {Phys. Rev. Lett.}\ }\textbf {\bibinfo {volume} {92}},\ \bibinfo {pages} {011101} (\bibinfo {year} {2004})},\ \Eprint {https://arxiv.org/abs/hep-ph/0307151} {arXiv:hep-ph/0307151} \BibitemShut {NoStop}%
\bibitem [{\citenamefont {Esmaili}(2010)}]{Esmaili:2009fk}%
  \BibitemOpen
  \bibfield  {author} {\bibinfo {author} {\bibfnamefont {A.}~\bibnamefont {Esmaili}},\ }\bibfield  {title} {\bibinfo {title} {{Pseudo-Dirac neutrino scenario: cosmic neutrinos at neutrino telescopes}},\ }\href {https://doi.org/10.1103/PhysRevD.81.013006} {\bibfield  {journal} {\bibinfo  {journal} {Phys. Rev. D}\ }\textbf {\bibinfo {volume} {81}},\ \bibinfo {pages} {013006} (\bibinfo {year} {2010})},\ \Eprint {https://arxiv.org/abs/0909.5410} {arXiv:0909.5410 [hep-ph]} \BibitemShut {NoStop}%
\bibitem [{\citenamefont {Carloni}\ \emph {et~al.}(2024)\citenamefont {Carloni}, \citenamefont {Mart{\'\i}nez-Soler}, \citenamefont {Arguelles}, \citenamefont {Babu},\ and\ \citenamefont {Dev}}]{Carloni:2022cqz}%
  \BibitemOpen
  \bibfield  {author} {\bibinfo {author} {\bibfnamefont {K.}~\bibnamefont {Carloni}}, \bibinfo {author} {\bibfnamefont {I.}~\bibnamefont {Mart{\'\i}nez-Soler}}, \bibinfo {author} {\bibfnamefont {C.~A.}\ \bibnamefont {Arguelles}}, \bibinfo {author} {\bibfnamefont {K.~S.}\ \bibnamefont {Babu}},\ and\ \bibinfo {author} {\bibfnamefont {P.~S.~B.}\ \bibnamefont {Dev}},\ }\bibfield  {title} {\bibinfo {title} {{Probing pseudo-Dirac neutrinos with astrophysical sources at IceCube}},\ }\href {https://doi.org/10.1103/PhysRevD.109.L051702} {\bibfield  {journal} {\bibinfo  {journal} {Phys. Rev. D}\ }\textbf {\bibinfo {volume} {109}},\ \bibinfo {pages} {L051702} (\bibinfo {year} {2024})},\ \Eprint {https://arxiv.org/abs/2212.00737} {arXiv:2212.00737 [astro-ph.HE]} \BibitemShut {NoStop}%
\bibitem [{\citenamefont {Blennow}\ and\ \citenamefont {Meloni}(2009)}]{Blennow:2009rp}%
  \BibitemOpen
  \bibfield  {author} {\bibinfo {author} {\bibfnamefont {M.}~\bibnamefont {Blennow}}\ and\ \bibinfo {author} {\bibfnamefont {D.}~\bibnamefont {Meloni}},\ }\bibfield  {title} {\bibinfo {title} {{Non-standard interaction effects on astrophysical neutrino fluxes}},\ }\href {https://doi.org/10.1103/PhysRevD.80.065009} {\bibfield  {journal} {\bibinfo  {journal} {Phys. Rev. D}\ }\textbf {\bibinfo {volume} {80}},\ \bibinfo {pages} {065009} (\bibinfo {year} {2009})},\ \Eprint {https://arxiv.org/abs/0901.2110} {arXiv:0901.2110 [hep-ph]} \BibitemShut {NoStop}%
\bibitem [{\citenamefont {Gonzalez-Garcia}\ \emph {et~al.}(2016)\citenamefont {Gonzalez-Garcia}, \citenamefont {Maltoni}, \citenamefont {Martinez-Soler},\ and\ \citenamefont {Song}}]{Gonzalez-Garcia:2016gpq}%
  \BibitemOpen
  \bibfield  {author} {\bibinfo {author} {\bibfnamefont {M.~C.}\ \bibnamefont {Gonzalez-Garcia}}, \bibinfo {author} {\bibfnamefont {M.}~\bibnamefont {Maltoni}}, \bibinfo {author} {\bibfnamefont {I.}~\bibnamefont {Martinez-Soler}},\ and\ \bibinfo {author} {\bibfnamefont {N.}~\bibnamefont {Song}},\ }\bibfield  {title} {\bibinfo {title} {{Non-standard neutrino interactions in the Earth and the flavor of astrophysical neutrinos}},\ }\href {https://doi.org/10.1016/j.astropartphys.2016.07.001} {\bibfield  {journal} {\bibinfo  {journal} {Astropart. Phys.}\ }\textbf {\bibinfo {volume} {84}},\ \bibinfo {pages} {15} (\bibinfo {year} {2016})},\ \Eprint {https://arxiv.org/abs/1605.08055} {arXiv:1605.08055 [hep-ph]} \BibitemShut {NoStop}%
\bibitem [{\citenamefont {de~Salas}\ \emph {et~al.}(2016)\citenamefont {de~Salas}, \citenamefont {Lineros},\ and\ \citenamefont {T{\'o}rtola}}]{deSalas:2016svi}%
  \BibitemOpen
  \bibfield  {author} {\bibinfo {author} {\bibfnamefont {P.~F.}\ \bibnamefont {de~Salas}}, \bibinfo {author} {\bibfnamefont {R.~A.}\ \bibnamefont {Lineros}},\ and\ \bibinfo {author} {\bibfnamefont {M.}~\bibnamefont {T{\'o}rtola}},\ }\bibfield  {title} {\bibinfo {title} {{Neutrino propagation in the galactic dark matter halo}},\ }\href {https://doi.org/10.1103/PhysRevD.94.123001} {\bibfield  {journal} {\bibinfo  {journal} {Phys. Rev. D}\ }\textbf {\bibinfo {volume} {94}},\ \bibinfo {pages} {123001} (\bibinfo {year} {2016})},\ \Eprint {https://arxiv.org/abs/1601.05798} {arXiv:1601.05798 [astro-ph.HE]} \BibitemShut {NoStop}%
\bibitem [{\citenamefont {Farzan}\ and\ \citenamefont {Palomares-Ruiz}(2019)}]{Farzan:2018pnk}%
  \BibitemOpen
  \bibfield  {author} {\bibinfo {author} {\bibfnamefont {Y.}~\bibnamefont {Farzan}}\ and\ \bibinfo {author} {\bibfnamefont {S.}~\bibnamefont {Palomares-Ruiz}},\ }\bibfield  {title} {\bibinfo {title} {{Flavor of cosmic neutrinos preserved by ultralight dark matter}},\ }\href {https://doi.org/10.1103/PhysRevD.99.051702} {\bibfield  {journal} {\bibinfo  {journal} {Phys. Rev. D}\ }\textbf {\bibinfo {volume} {99}},\ \bibinfo {pages} {051702} (\bibinfo {year} {2019})},\ \Eprint {https://arxiv.org/abs/1810.00892} {arXiv:1810.00892 [hep-ph]} \BibitemShut {NoStop}%
\bibitem [{\citenamefont {Reynoso}\ \emph {et~al.}(2022)\citenamefont {Reynoso}, \citenamefont {Sampayo},\ and\ \citenamefont {Carulli}}]{Reynoso:2022vrn}%
  \BibitemOpen
  \bibfield  {author} {\bibinfo {author} {\bibfnamefont {M.~M.}\ \bibnamefont {Reynoso}}, \bibinfo {author} {\bibfnamefont {O.~A.}\ \bibnamefont {Sampayo}},\ and\ \bibinfo {author} {\bibfnamefont {A.~M.}\ \bibnamefont {Carulli}},\ }\bibfield  {title} {\bibinfo {title} {{Neutrino interactions with ultralight axion-like dark matter}},\ }\href {https://doi.org/10.1140/epjc/s10052-022-10228-w} {\bibfield  {journal} {\bibinfo  {journal} {Eur. Phys. J. C}\ }\textbf {\bibinfo {volume} {82}},\ \bibinfo {pages} {274} (\bibinfo {year} {2022})},\ \Eprint {https://arxiv.org/abs/2203.11642} {arXiv:2203.11642 [hep-ph]} \BibitemShut {NoStop}%
\bibitem [{\citenamefont {Arg\"uelles}\ \emph {et~al.}(2023{\natexlab{b}})\citenamefont {Arg\"uelles}, \citenamefont {Farrag},\ and\ \citenamefont {Katori}}]{Arguelles:2023wvf}%
  \BibitemOpen
  \bibfield  {author} {\bibinfo {author} {\bibfnamefont {C.~A.}\ \bibnamefont {Arg\"uelles}}, \bibinfo {author} {\bibfnamefont {K.}~\bibnamefont {Farrag}},\ and\ \bibinfo {author} {\bibfnamefont {T.}~\bibnamefont {Katori}},\ }\bibfield  {title} {\bibinfo {title} {{Ultra-light dark matter limits from astrophysical neutrino flavour}},\ }\href {https://doi.org/10.22323/1.444.1415} {\bibfield  {journal} {\bibinfo  {journal} {PoS}\ }\textbf {\bibinfo {volume} {ICRC2023}},\ \bibinfo {pages} {1415} (\bibinfo {year} {2023}{\natexlab{b}})}\BibitemShut {NoStop}%
\bibitem [{\citenamefont {Hung}\ and\ \citenamefont {Pas}(2005)}]{Hung:2003jb}%
  \BibitemOpen
  \bibfield  {author} {\bibinfo {author} {\bibfnamefont {P.~Q.}\ \bibnamefont {Hung}}\ and\ \bibinfo {author} {\bibfnamefont {H.}~\bibnamefont {Pas}},\ }\bibfield  {title} {\bibinfo {title} {{Cosmo MSW effect for mass varying neutrinos}},\ }\href {https://doi.org/10.1142/S0217732305016981} {\bibfield  {journal} {\bibinfo  {journal} {Mod. Phys. Lett. A}\ }\textbf {\bibinfo {volume} {20}},\ \bibinfo {pages} {1209} (\bibinfo {year} {2005})},\ \Eprint {https://arxiv.org/abs/astro-ph/0311131} {arXiv:astro-ph/0311131} \BibitemShut {NoStop}%
\bibitem [{\citenamefont {Ando}\ \emph {et~al.}(2009)\citenamefont {Ando}, \citenamefont {Kamionkowski},\ and\ \citenamefont {Mocioiu}}]{Ando:2009ts}%
  \BibitemOpen
  \bibfield  {author} {\bibinfo {author} {\bibfnamefont {S.}~\bibnamefont {Ando}}, \bibinfo {author} {\bibfnamefont {M.}~\bibnamefont {Kamionkowski}},\ and\ \bibinfo {author} {\bibfnamefont {I.}~\bibnamefont {Mocioiu}},\ }\bibfield  {title} {\bibinfo {title} {{Neutrino oscillations, Lorentz/CPT violation, and dark energy}},\ }\href {https://doi.org/10.1103/PhysRevD.80.123522} {\bibfield  {journal} {\bibinfo  {journal} {Phys. Rev. D}\ }\textbf {\bibinfo {volume} {80}},\ \bibinfo {pages} {123522} (\bibinfo {year} {2009})},\ \Eprint {https://arxiv.org/abs/0910.4391} {arXiv:0910.4391 [hep-ph]} \BibitemShut {NoStop}%
\bibitem [{\citenamefont {Klop}\ and\ \citenamefont {Ando}(2018)}]{Klop:2017dim}%
  \BibitemOpen
  \bibfield  {author} {\bibinfo {author} {\bibfnamefont {N.}~\bibnamefont {Klop}}\ and\ \bibinfo {author} {\bibfnamefont {S.}~\bibnamefont {Ando}},\ }\bibfield  {title} {\bibinfo {title} {{Effects of a neutrino-dark energy coupling on oscillations of high-energy neutrinos}},\ }\href {https://doi.org/10.1103/PhysRevD.97.063006} {\bibfield  {journal} {\bibinfo  {journal} {Phys. Rev. D}\ }\textbf {\bibinfo {volume} {97}},\ \bibinfo {pages} {063006} (\bibinfo {year} {2018})},\ \Eprint {https://arxiv.org/abs/1712.05413} {arXiv:1712.05413 [hep-ph]} \BibitemShut {NoStop}%
\bibitem [{\citenamefont {Athar}\ \emph {et~al.}(2000{\natexlab{b}})\citenamefont {Athar}, \citenamefont {Jezabek},\ and\ \citenamefont {Yasuda}}]{Athar:2000yw}%
  \BibitemOpen
  \bibfield  {author} {\bibinfo {author} {\bibfnamefont {H.}~\bibnamefont {Athar}}, \bibinfo {author} {\bibfnamefont {M.}~\bibnamefont {Jezabek}},\ and\ \bibinfo {author} {\bibfnamefont {O.}~\bibnamefont {Yasuda}},\ }\bibfield  {title} {\bibinfo {title} {{Effects of neutrino mixing on high-energy cosmic neutrino flux}},\ }\href {https://doi.org/10.1103/PhysRevD.62.103007} {\bibfield  {journal} {\bibinfo  {journal} {Phys. Rev. D}\ }\textbf {\bibinfo {volume} {62}},\ \bibinfo {pages} {103007} (\bibinfo {year} {2000}{\natexlab{b}})},\ \Eprint {https://arxiv.org/abs/hep-ph/0005104} {arXiv:hep-ph/0005104} \BibitemShut {NoStop}%
\bibitem [{\citenamefont {Keranen}\ \emph {et~al.}(2003)\citenamefont {Keranen}, \citenamefont {Maalampi}, \citenamefont {Myyrylainen},\ and\ \citenamefont {Riittinen}}]{Keranen:2003xd}%
  \BibitemOpen
  \bibfield  {author} {\bibinfo {author} {\bibfnamefont {P.}~\bibnamefont {Keranen}}, \bibinfo {author} {\bibfnamefont {J.}~\bibnamefont {Maalampi}}, \bibinfo {author} {\bibfnamefont {M.}~\bibnamefont {Myyrylainen}},\ and\ \bibinfo {author} {\bibfnamefont {J.}~\bibnamefont {Riittinen}},\ }\bibfield  {title} {\bibinfo {title} {{Effects of sterile neutrinos on the ultrahigh-energy cosmic neutrino flux}},\ }\href {https://doi.org/10.1016/j.physletb.2003.09.006} {\bibfield  {journal} {\bibinfo  {journal} {Phys. Lett. B}\ }\textbf {\bibinfo {volume} {574}},\ \bibinfo {pages} {162} (\bibinfo {year} {2003})},\ \Eprint {https://arxiv.org/abs/hep-ph/0307041} {arXiv:hep-ph/0307041} \BibitemShut {NoStop}%
\bibitem [{\citenamefont {Arg\"uelles}\ \emph {et~al.}(2020{\natexlab{b}})\citenamefont {Arg\"uelles}, \citenamefont {Farrag}, \citenamefont {Katori}, \citenamefont {Khandelwal}, \citenamefont {Mandalia},\ and\ \citenamefont {Salvado}}]{Arguelles:2019tum}%
  \BibitemOpen
  \bibfield  {author} {\bibinfo {author} {\bibfnamefont {C.~A.}\ \bibnamefont {Arg\"uelles}}, \bibinfo {author} {\bibfnamefont {K.}~\bibnamefont {Farrag}}, \bibinfo {author} {\bibfnamefont {T.}~\bibnamefont {Katori}}, \bibinfo {author} {\bibfnamefont {R.}~\bibnamefont {Khandelwal}}, \bibinfo {author} {\bibfnamefont {S.}~\bibnamefont {Mandalia}},\ and\ \bibinfo {author} {\bibfnamefont {J.}~\bibnamefont {Salvado}},\ }\bibfield  {title} {\bibinfo {title} {{Sterile neutrinos in astrophysical neutrino flavor}},\ }\href {https://doi.org/10.1088/1475-7516/2020/02/015} {\bibfield  {journal} {\bibinfo  {journal} {JCAP}\ }\textbf {\bibinfo {volume} {02}},\ \bibinfo {pages} {015}},\ \Eprint {https://arxiv.org/abs/1909.05341} {arXiv:1909.05341 [hep-ph]} \BibitemShut {NoStop}%
\bibitem [{\citenamefont {Barenboim}\ and\ \citenamefont {Quigg}(2003)}]{Barenboim:2003jm}%
  \BibitemOpen
  \bibfield  {author} {\bibinfo {author} {\bibfnamefont {G.}~\bibnamefont {Barenboim}}\ and\ \bibinfo {author} {\bibfnamefont {C.}~\bibnamefont {Quigg}},\ }\bibfield  {title} {\bibinfo {title} {{Neutrino observatories can characterize cosmic sources and neutrino properties}},\ }\href {https://doi.org/10.1103/PhysRevD.67.073024} {\bibfield  {journal} {\bibinfo  {journal} {Phys. Rev. D}\ }\textbf {\bibinfo {volume} {67}},\ \bibinfo {pages} {073024} (\bibinfo {year} {2003})},\ \Eprint {https://arxiv.org/abs/hep-ph/0301220} {arXiv:hep-ph/0301220} \BibitemShut {NoStop}%
\bibitem [{\citenamefont {Hooper}\ \emph {et~al.}(2005{\natexlab{a}})\citenamefont {Hooper}, \citenamefont {Morgan},\ and\ \citenamefont {Winstanley}}]{Hooper:2005jp}%
  \BibitemOpen
  \bibfield  {author} {\bibinfo {author} {\bibfnamefont {D.}~\bibnamefont {Hooper}}, \bibinfo {author} {\bibfnamefont {D.}~\bibnamefont {Morgan}},\ and\ \bibinfo {author} {\bibfnamefont {E.}~\bibnamefont {Winstanley}},\ }\bibfield  {title} {\bibinfo {title} {{Lorentz and CPT invariance violation in high-energy neutrinos}},\ }\href {https://doi.org/10.1103/PhysRevD.72.065009} {\bibfield  {journal} {\bibinfo  {journal} {Phys. Rev. D}\ }\textbf {\bibinfo {volume} {72}},\ \bibinfo {pages} {065009} (\bibinfo {year} {2005}{\natexlab{a}})},\ \Eprint {https://arxiv.org/abs/hep-ph/0506091} {arXiv:hep-ph/0506091} \BibitemShut {NoStop}%
\bibitem [{\citenamefont {Bhattacharya}\ \emph {et~al.}(2010{\natexlab{b}})\citenamefont {Bhattacharya}, \citenamefont {Choubey}, \citenamefont {Gandhi},\ and\ \citenamefont {Watanabe}}]{Bhattacharya:2009tx}%
  \BibitemOpen
  \bibfield  {author} {\bibinfo {author} {\bibfnamefont {A.}~\bibnamefont {Bhattacharya}}, \bibinfo {author} {\bibfnamefont {S.}~\bibnamefont {Choubey}}, \bibinfo {author} {\bibfnamefont {R.}~\bibnamefont {Gandhi}},\ and\ \bibinfo {author} {\bibfnamefont {A.}~\bibnamefont {Watanabe}},\ }\bibfield  {title} {\bibinfo {title} {{Diffuse ultra-high energy neutrino fluxes and physics beyond the Standard Model}},\ }\href {https://doi.org/10.1016/j.physletb.2010.04.078} {\bibfield  {journal} {\bibinfo  {journal} {Phys. Lett. B}\ }\textbf {\bibinfo {volume} {690}},\ \bibinfo {pages} {42} (\bibinfo {year} {2010}{\natexlab{b}})},\ \Eprint {https://arxiv.org/abs/0910.4396} {arXiv:0910.4396 [hep-ph]} \BibitemShut {NoStop}%
\bibitem [{\citenamefont {Bustamante}\ \emph {et~al.}(2010)\citenamefont {Bustamante}, \citenamefont {Gago},\ and\ \citenamefont {Pena-Garay}}]{Bustamante:2010nq}%
  \BibitemOpen
  \bibfield  {author} {\bibinfo {author} {\bibfnamefont {M.}~\bibnamefont {Bustamante}}, \bibinfo {author} {\bibfnamefont {A.~M.}\ \bibnamefont {Gago}},\ and\ \bibinfo {author} {\bibfnamefont {C.}~\bibnamefont {Pena-Garay}},\ }\bibfield  {title} {\bibinfo {title} {{Energy-independent new physics in the flavour ratios of high-energy astrophysical neutrinos}},\ }\href {https://doi.org/10.1007/JHEP04(2010)066} {\bibfield  {journal} {\bibinfo  {journal} {JHEP}\ }\textbf {\bibinfo {volume} {04}},\ \bibinfo {pages} {066}},\ \Eprint {https://arxiv.org/abs/1001.4878} {arXiv:1001.4878 [hep-ph]} \BibitemShut {NoStop}%
\bibitem [{\citenamefont {Katori}\ \emph {et~al.}(2017)\citenamefont {Katori}, \citenamefont {Arg{\"u}elles},\ and\ \citenamefont {Salvado}}]{Katori:2016eni}%
  \BibitemOpen
  \bibfield  {author} {\bibinfo {author} {\bibfnamefont {T.}~\bibnamefont {Katori}}, \bibinfo {author} {\bibfnamefont {C.~A.}\ \bibnamefont {Arg{\"u}elles}},\ and\ \bibinfo {author} {\bibfnamefont {J.}~\bibnamefont {Salvado}},\ }\bibfield  {title} {\bibinfo {title} {{Test of Lorentz violation with astrophysical neutrino flavor}},\ }in\ \href {https://doi.org/10.1142/9789813148505_0053} {\emph {\bibinfo {booktitle} {{7th Meeting on CPT and Lorentz Symmetry}}}}\ (\bibinfo {year} {2017})\ pp.\ \bibinfo {pages} {209--212},\ \Eprint {https://arxiv.org/abs/1607.08448} {arXiv:1607.08448 [hep-ph]} \BibitemShut {NoStop}%
\bibitem [{\citenamefont {Minakata}\ and\ \citenamefont {Smirnov}(1996)}]{Minakata:1996nd}%
  \BibitemOpen
  \bibfield  {author} {\bibinfo {author} {\bibfnamefont {H.}~\bibnamefont {Minakata}}\ and\ \bibinfo {author} {\bibfnamefont {A.~Y.}\ \bibnamefont {Smirnov}},\ }\bibfield  {title} {\bibinfo {title} {{High-energy cosmic neutrinos and the equivalence principle}},\ }\href {https://doi.org/10.1103/PhysRevD.54.3698} {\bibfield  {journal} {\bibinfo  {journal} {Phys. Rev. D}\ }\textbf {\bibinfo {volume} {54}},\ \bibinfo {pages} {3698} (\bibinfo {year} {1996})},\ \Eprint {https://arxiv.org/abs/hep-ph/9601311} {arXiv:hep-ph/9601311} \BibitemShut {NoStop}%
\bibitem [{\citenamefont {Hooper}\ \emph {et~al.}(2005{\natexlab{b}})\citenamefont {Hooper}, \citenamefont {Morgan},\ and\ \citenamefont {Winstanley}}]{Hooper:2004xr}%
  \BibitemOpen
  \bibfield  {author} {\bibinfo {author} {\bibfnamefont {D.}~\bibnamefont {Hooper}}, \bibinfo {author} {\bibfnamefont {D.}~\bibnamefont {Morgan}},\ and\ \bibinfo {author} {\bibfnamefont {E.}~\bibnamefont {Winstanley}},\ }\bibfield  {title} {\bibinfo {title} {{Probing quantum decoherence with high-energy neutrinos}},\ }\href {https://doi.org/10.1016/j.physletb.2005.01.034} {\bibfield  {journal} {\bibinfo  {journal} {Phys. Lett. B}\ }\textbf {\bibinfo {volume} {609}},\ \bibinfo {pages} {206} (\bibinfo {year} {2005}{\natexlab{b}})},\ \Eprint {https://arxiv.org/abs/hep-ph/0410094} {arXiv:hep-ph/0410094} \BibitemShut {NoStop}%
\bibitem [{\citenamefont {Anchordoqui}\ \emph {et~al.}(2005)\citenamefont {Anchordoqui}, \citenamefont {Goldberg}, \citenamefont {Gonzalez-Garcia}, \citenamefont {Halzen}, \citenamefont {Hooper}, \citenamefont {Sarkar},\ and\ \citenamefont {Weiler}}]{Anchordoqui:2005gj}%
  \BibitemOpen
  \bibfield  {author} {\bibinfo {author} {\bibfnamefont {L.~A.}\ \bibnamefont {Anchordoqui}}, \bibinfo {author} {\bibfnamefont {H.}~\bibnamefont {Goldberg}}, \bibinfo {author} {\bibfnamefont {M.~C.}\ \bibnamefont {Gonzalez-Garcia}}, \bibinfo {author} {\bibfnamefont {F.}~\bibnamefont {Halzen}}, \bibinfo {author} {\bibfnamefont {D.}~\bibnamefont {Hooper}}, \bibinfo {author} {\bibfnamefont {S.}~\bibnamefont {Sarkar}},\ and\ \bibinfo {author} {\bibfnamefont {T.~J.}\ \bibnamefont {Weiler}},\ }\bibfield  {title} {\bibinfo {title} {{Probing Planck scale physics with IceCube}},\ }\href {https://doi.org/10.1103/PhysRevD.72.065019} {\bibfield  {journal} {\bibinfo  {journal} {Phys. Rev. D}\ }\textbf {\bibinfo {volume} {72}},\ \bibinfo {pages} {065019} (\bibinfo {year} {2005})},\ \Eprint {https://arxiv.org/abs/hep-ph/0506168} {arXiv:hep-ph/0506168} \BibitemShut {NoStop}%
\bibitem [{\citenamefont {Learned}(1981)}]{Learned1981DUMANDTau}%
  \BibitemOpen
  \bibfield  {author} {\bibinfo {author} {\bibfnamefont {J.~G.}\ \bibnamefont {Learned}},\ }\bibfield  {title} {\bibinfo {title} {Dumand as a tau detector},\ }in\ \href@noop {} {\emph {\bibinfo {booktitle} {Proceedings of the 1980 International DUMAND Symposium}}},\ Vol.~\bibinfo {volume} {2},\ \bibinfo {editor} {edited by\ \bibinfo {editor} {\bibfnamefont {V.~J.}\ \bibnamefont {Stenger}}}\ (\bibinfo  {publisher} {University of Hawaii / Hawaii DUMAND Center},\ \bibinfo {address} {Honolulu, Hawaii},\ \bibinfo {year} {1981})\ p.\ \bibinfo {pages} {272}\BibitemShut {NoStop}%
\bibitem [{\citenamefont {Cowen}(2007)}]{Cowen:2007ny}%
  \BibitemOpen
  \bibfield  {author} {\bibinfo {author} {\bibfnamefont {D.~F.}\ \bibnamefont {Cowen}} (\bibinfo {collaboration} {IceCube}),\ }\bibfield  {title} {\bibinfo {title} {{Tau neutrinos in IceCube}},\ }\href {https://doi.org/10.1088/1742-6596/60/1/048} {\bibfield  {journal} {\bibinfo  {journal} {J. Phys. Conf. Ser.}\ }\textbf {\bibinfo {volume} {60}},\ \bibinfo {pages} {227} (\bibinfo {year} {2007})}\BibitemShut {NoStop}%
\bibitem [{\citenamefont {Beacom}\ \emph {et~al.}(2003{\natexlab{b}})\citenamefont {Beacom}, \citenamefont {Bell}, \citenamefont {Hooper}, \citenamefont {Pakvasa},\ and\ \citenamefont {Weiler}}]{Beacom:2003nh}%
  \BibitemOpen
  \bibfield  {author} {\bibinfo {author} {\bibfnamefont {J.~F.}\ \bibnamefont {Beacom}}, \bibinfo {author} {\bibfnamefont {N.~F.}\ \bibnamefont {Bell}}, \bibinfo {author} {\bibfnamefont {D.}~\bibnamefont {Hooper}}, \bibinfo {author} {\bibfnamefont {S.}~\bibnamefont {Pakvasa}},\ and\ \bibinfo {author} {\bibfnamefont {T.~J.}\ \bibnamefont {Weiler}},\ }\bibfield  {title} {\bibinfo {title} {{Measuring flavor ratios of high-energy astrophysical neutrinos}},\ }\href {https://doi.org/10.1103/PhysRevD.68.093005} {\bibfield  {journal} {\bibinfo  {journal} {Phys. Rev. D}\ }\textbf {\bibinfo {volume} {68}},\ \bibinfo {pages} {093005} (\bibinfo {year} {2003}{\natexlab{b}})},\ \bibinfo {note} {[Erratum: Phys.Rev.D 72, 019901 (2005)]},\ \Eprint {https://arxiv.org/abs/hep-ph/0307025} {arXiv:hep-ph/0307025} \BibitemShut {NoStop}%
\bibitem [{\citenamefont {Bugaev}\ \emph {et~al.}(2004)\citenamefont {Bugaev}, \citenamefont {Montaruli}, \citenamefont {Shlepin},\ and\ \citenamefont {Sokalski}}]{Bugaev:2003sw}%
  \BibitemOpen
  \bibfield  {author} {\bibinfo {author} {\bibfnamefont {E.}~\bibnamefont {Bugaev}}, \bibinfo {author} {\bibfnamefont {T.}~\bibnamefont {Montaruli}}, \bibinfo {author} {\bibfnamefont {Y.}~\bibnamefont {Shlepin}},\ and\ \bibinfo {author} {\bibfnamefont {I.~A.}\ \bibnamefont {Sokalski}},\ }\bibfield  {title} {\bibinfo {title} {{Propagation of tau neutrinos and tau leptons through the earth and their detection in underwater / ice neutrino telescopes}},\ }\href {https://doi.org/10.1016/j.astropartphys.2004.03.002} {\bibfield  {journal} {\bibinfo  {journal} {Astropart. Phys.}\ }\textbf {\bibinfo {volume} {21}},\ \bibinfo {pages} {491} (\bibinfo {year} {2004})},\ \Eprint {https://arxiv.org/abs/hep-ph/0312295} {arXiv:hep-ph/0312295} \BibitemShut {NoStop}%
\bibitem [{\citenamefont {DeYoung}\ \emph {et~al.}(2007)\citenamefont {DeYoung}, \citenamefont {Razzaque},\ and\ \citenamefont {Cowen}}]{DeYoung:2006fg}%
  \BibitemOpen
  \bibfield  {author} {\bibinfo {author} {\bibfnamefont {T.}~\bibnamefont {DeYoung}}, \bibinfo {author} {\bibfnamefont {S.}~\bibnamefont {Razzaque}},\ and\ \bibinfo {author} {\bibfnamefont {D.~F.}\ \bibnamefont {Cowen}},\ }\bibfield  {title} {\bibinfo {title} {{Astrophysical tau neutrino detection in kilometer-scale Cherenkov detectors via muonic tau decay}},\ }\href {https://doi.org/10.1016/j.astropartphys.2006.11.003} {\bibfield  {journal} {\bibinfo  {journal} {Astropart. Phys.}\ }\textbf {\bibinfo {volume} {27}},\ \bibinfo {pages} {238} (\bibinfo {year} {2007})},\ \Eprint {https://arxiv.org/abs/astro-ph/0608486} {arXiv:astro-ph/0608486} \BibitemShut {NoStop}%
\bibitem [{\citenamefont {Li}\ \emph {et~al.}(2019)\citenamefont {Li}, \citenamefont {Bustamante},\ and\ \citenamefont {Beacom}}]{Li:2016kra}%
  \BibitemOpen
  \bibfield  {author} {\bibinfo {author} {\bibfnamefont {S.~W.}\ \bibnamefont {Li}}, \bibinfo {author} {\bibfnamefont {M.}~\bibnamefont {Bustamante}},\ and\ \bibinfo {author} {\bibfnamefont {J.~F.}\ \bibnamefont {Beacom}},\ }\bibfield  {title} {\bibinfo {title} {{Echo technique to distinguish flavors of astrophysical neutrinos}},\ }\href {https://doi.org/10.1103/PhysRevLett.122.151101} {\bibfield  {journal} {\bibinfo  {journal} {Phys. Rev. Lett.}\ }\textbf {\bibinfo {volume} {122}},\ \bibinfo {pages} {151101} (\bibinfo {year} {2019})},\ \Eprint {https://arxiv.org/abs/1606.06290} {arXiv:1606.06290 [astro-ph.HE]} \BibitemShut {NoStop}%
\bibitem [{\citenamefont {Kistler}\ and\ \citenamefont {Laha}(2018)}]{Kistler:2016ask}%
  \BibitemOpen
  \bibfield  {author} {\bibinfo {author} {\bibfnamefont {M.~D.}\ \bibnamefont {Kistler}}\ and\ \bibinfo {author} {\bibfnamefont {R.}~\bibnamefont {Laha}},\ }\bibfield  {title} {\bibinfo {title} {{Multi-PeV signals from a new astrophysical neutrino flux beyond the Glashow resonance}},\ }\href {https://doi.org/10.1103/PhysRevLett.120.241105} {\bibfield  {journal} {\bibinfo  {journal} {Phys. Rev. Lett.}\ }\textbf {\bibinfo {volume} {120}},\ \bibinfo {pages} {241105} (\bibinfo {year} {2018})},\ \Eprint {https://arxiv.org/abs/1605.08781} {arXiv:1605.08781 [astro-ph.HE]} \BibitemShut {NoStop}%
\bibitem [{\citenamefont {Abbasi}\ \emph {et~al.}(2022{\natexlab{a}})\citenamefont {Abbasi} \emph {et~al.}}]{IceCube:2020fpi}%
  \BibitemOpen
  \bibfield  {author} {\bibinfo {author} {\bibfnamefont {R.}~\bibnamefont {Abbasi}} \emph {et~al.} (\bibinfo {collaboration} {IceCube Collaboration}),\ }\bibfield  {title} {\bibinfo {title} {{Detection of astrophysical tau neutrino candidates in IceCube}},\ }\href {https://doi.org/10.1140/epjc/s10052-022-10795-y} {\bibfield  {journal} {\bibinfo  {journal} {Eur. Phys. J. C}\ }\textbf {\bibinfo {volume} {82}},\ \bibinfo {pages} {1031} (\bibinfo {year} {2022}{\natexlab{a}})},\ \Eprint {https://arxiv.org/abs/2011.03561} {arXiv:2011.03561 [hep-ex]} \BibitemShut {NoStop}%
\bibitem [{\citenamefont {Abbasi}\ \emph {et~al.}(2024{\natexlab{a}})\citenamefont {Abbasi} \emph {et~al.}}]{IceCube:2024nhk}%
  \BibitemOpen
  \bibfield  {author} {\bibinfo {author} {\bibfnamefont {R.}~\bibnamefont {Abbasi}} \emph {et~al.} (\bibinfo {collaboration} {IceCube Collaboration}),\ }\bibfield  {title} {\bibinfo {title} {{Observation of seven astrophysical tau neutrino candidates with IceCube}},\ }\href {https://doi.org/10.1103/PhysRevLett.132.151001} {\bibfield  {journal} {\bibinfo  {journal} {Phys. Rev. Lett.}\ }\textbf {\bibinfo {volume} {132}},\ \bibinfo {pages} {151001} (\bibinfo {year} {2024}{\natexlab{a}})},\ \Eprint {https://arxiv.org/abs/2403.02516} {arXiv:2403.02516 [astro-ph.HE]} \BibitemShut {NoStop}%
\bibitem [{\citenamefont {Aartsen}\ \emph {et~al.}(2017)\citenamefont {Aartsen} \emph {et~al.}}]{IceCube:2016zyt}%
  \BibitemOpen
  \bibfield  {author} {\bibinfo {author} {\bibfnamefont {M.~G.}\ \bibnamefont {Aartsen}} \emph {et~al.} (\bibinfo {collaboration} {IceCube Collaboration}),\ }\bibfield  {title} {\bibinfo {title} {{The IceCube neutrino observatory: instrumentation and online systems}},\ }\href {https://doi.org/10.1088/1748-0221/12/03/P03012} {\bibfield  {journal} {\bibinfo  {journal} {JINST}\ }\textbf {\bibinfo {volume} {12}}\bibfield  {number} {\bibinfo  {number} { (03)},\ \bibinfo {pages} {P03012}},\ }\bibinfo {note} {[Erratum: JINST 19, E05001 (2024)]},\ \Eprint {https://arxiv.org/abs/1612.05093} {arXiv:1612.05093 [astro-ph.IM]} \BibitemShut {NoStop}%
\bibitem [{\citenamefont {Adrian-Martinez}\ \emph {et~al.}(2016)\citenamefont {Adrian-Martinez} \emph {et~al.}}]{KM3Net:2016zxf}%
  \BibitemOpen
  \bibfield  {author} {\bibinfo {author} {\bibfnamefont {S.}~\bibnamefont {Adrian-Martinez}} \emph {et~al.} (\bibinfo {collaboration} {KM3NeT Collaboration}),\ }\bibfield  {title} {\bibinfo {title} {{Letter of intent for KM3NeT 2.0}},\ }\href {https://doi.org/10.1088/0954-3899/43/8/084001} {\bibfield  {journal} {\bibinfo  {journal} {J. Phys. G}\ }\textbf {\bibinfo {volume} {43}},\ \bibinfo {pages} {084001} (\bibinfo {year} {2016})},\ \Eprint {https://arxiv.org/abs/1601.07459} {arXiv:1601.07459 [astro-ph.IM]} \BibitemShut {NoStop}%
\bibitem [{\citenamefont {Agostini}\ \emph {et~al.}(2020)\citenamefont {Agostini} \emph {et~al.}}]{P-ONE:2020ljt}%
  \BibitemOpen
  \bibfield  {author} {\bibinfo {author} {\bibfnamefont {M.}~\bibnamefont {Agostini}} \emph {et~al.} (\bibinfo {collaboration} {P-ONE Collaboration}),\ }\bibfield  {title} {\bibinfo {title} {{The Pacific Ocean Neutrino Experiment}},\ }\href {https://doi.org/10.1038/s41550-020-1182-4} {\bibfield  {journal} {\bibinfo  {journal} {Nature Astron.}\ }\textbf {\bibinfo {volume} {4}},\ \bibinfo {pages} {913} (\bibinfo {year} {2020})},\ \Eprint {https://arxiv.org/abs/2005.09493} {arXiv:2005.09493 [astro-ph.HE]} \BibitemShut {NoStop}%
\bibitem [{\citenamefont {Avrorin}\ \emph {et~al.}(2015)\citenamefont {Avrorin} \emph {et~al.}}]{Avrorin:2015wba}%
  \BibitemOpen
  \bibfield  {author} {\bibinfo {author} {\bibfnamefont {A.~D.}\ \bibnamefont {Avrorin}} \emph {et~al.},\ }\bibfield  {title} {\bibinfo {title} {{Status and recent results of the BAIKAL-GVD project}},\ }\href {https://doi.org/10.1134/S1063779615020033} {\bibfield  {journal} {\bibinfo  {journal} {Phys. Part. Nucl.}\ }\textbf {\bibinfo {volume} {46}},\ \bibinfo {pages} {211} (\bibinfo {year} {2015})}\BibitemShut {NoStop}%
\bibitem [{\citenamefont {Ye}\ \emph {et~al.}(2023)\citenamefont {Ye} \emph {et~al.}}]{TRIDENT:2022hql}%
  \BibitemOpen
  \bibfield  {author} {\bibinfo {author} {\bibfnamefont {Z.~P.}\ \bibnamefont {Ye}} \emph {et~al.} (\bibinfo {collaboration} {TRIDENT Collaboration}),\ }\bibfield  {title} {\bibinfo {title} {{A multi-cubic-kilometre neutrino telescope in the western Pacific Ocean}},\ }\href {https://doi.org/10.1038/s41550-023-02087-6} {\bibfield  {journal} {\bibinfo  {journal} {Nature Astron.}\ }\textbf {\bibinfo {volume} {7}},\ \bibinfo {pages} {1497} (\bibinfo {year} {2023})},\ \Eprint {https://arxiv.org/abs/2207.04519} {arXiv:2207.04519 [astro-ph.HE]} \BibitemShut {NoStop}%
\bibitem [{\citenamefont {Arg{\"u}elles}\ \emph {et~al.}(2025)\citenamefont {Arg{\"u}elles} \emph {et~al.}}]{TAMBO:2025jio}%
  \BibitemOpen
  \bibfield  {author} {\bibinfo {author} {\bibfnamefont {C.~A.}\ \bibnamefont {Arg{\"u}elles}} \emph {et~al.} (\bibinfo {collaboration} {TAMBO Collaboration}),\ }\bibfield  {title} {\bibinfo {title} {{TAMBO: A deep-valley neutrino observatory}},\ }\href@noop {} {\  (\bibinfo {year} {2025})},\ \Eprint {https://arxiv.org/abs/2507.08070} {arXiv:2507.08070 [astro-ph.HE]} \BibitemShut {NoStop}%
\bibitem [{\citenamefont {Otte}(2025)}]{Otte:2025dld}%
  \BibitemOpen
  \bibfield  {author} {\bibinfo {author} {\bibfnamefont {A.~N.}\ \bibnamefont {Otte}} (\bibinfo {collaboration} {Trinity Collaboration}),\ }\bibfield  {title} {\bibinfo {title} {{The Trinity-one PeV-neutrino telescope}},\ }\href {https://doi.org/10.22323/1.501.1136} {\bibfield  {journal} {\bibinfo  {journal} {PoS}\ }\textbf {\bibinfo {volume} {ICRC2025}},\ \bibinfo {pages} {1136} (\bibinfo {year} {2025})},\ \Eprint {https://arxiv.org/abs/2509.18237} {arXiv:2509.18237 [astro-ph.HE]} \BibitemShut {NoStop}%
\bibitem [{\citenamefont {{\'A}lvarez-Mu{\~n}iz}\ \emph {et~al.}(2020)\citenamefont {{\'A}lvarez-Mu{\~n}iz} \emph {et~al.}}]{GRAND:2018iaj}%
  \BibitemOpen
  \bibfield  {author} {\bibinfo {author} {\bibfnamefont {J.}~\bibnamefont {{\'A}lvarez-Mu{\~n}iz}} \emph {et~al.} (\bibinfo {collaboration} {GRAND Collaboration}),\ }\bibfield  {title} {\bibinfo {title} {{The Giant Radio Array for Neutrino Detection (GRAND): science and design}},\ }\href {https://doi.org/10.1007/s11433-018-9385-7} {\bibfield  {journal} {\bibinfo  {journal} {Sci. China Phys. Mech. Astron.}\ }\textbf {\bibinfo {volume} {63}},\ \bibinfo {pages} {219501} (\bibinfo {year} {2020})},\ \Eprint {https://arxiv.org/abs/1810.09994} {arXiv:1810.09994 [astro-ph.HE]} \BibitemShut {NoStop}%
\bibitem [{\citenamefont {Kotera}\ \emph {et~al.}(2025)\citenamefont {Kotera} \emph {et~al.}}]{GRAND:2025rps}%
  \BibitemOpen
  \bibfield  {author} {\bibinfo {author} {\bibfnamefont {K.}~\bibnamefont {Kotera}} \emph {et~al.} (\bibinfo {collaboration} {BEACON and GRAND Collaborations}),\ }\bibfield  {title} {\bibinfo {title} {{The Hybrid Elevated Radio Observatory for Neutrinos (HERON) project}},\ }\href {https://doi.org/10.22323/1.501.1078} {\bibfield  {journal} {\bibinfo  {journal} {PoS}\ }\textbf {\bibinfo {volume} {ICRC2025}},\ \bibinfo {pages} {1078} (\bibinfo {year} {2025})},\ \Eprint {https://arxiv.org/abs/2507.04382} {arXiv:2507.04382 [astro-ph.IM]} \BibitemShut {NoStop}%
\bibitem [{\citenamefont {Aguilar}\ \emph {et~al.}(2021)\citenamefont {Aguilar} \emph {et~al.}}]{RNO-G:2020rmc}%
  \BibitemOpen
  \bibfield  {author} {\bibinfo {author} {\bibfnamefont {J.~A.}\ \bibnamefont {Aguilar}} \emph {et~al.} (\bibinfo {collaboration} {RNO-G Collaboration}),\ }\bibfield  {title} {\bibinfo {title} {{Design and sensitivity of the Radio Neutrino Observatory in Greenland (RNO-G)}},\ }\href {https://doi.org/10.1088/1748-0221/16/03/P03025} {\bibfield  {journal} {\bibinfo  {journal} {JINST}\ }\textbf {\bibinfo {volume} {16}}\bibfield  {number} {\bibinfo  {number} { (03)},\ \bibinfo {pages} {P03025}},\ }\bibinfo {note} {[Erratum: JINST 18, E03001 (2023)]},\ \Eprint {https://arxiv.org/abs/2010.12279} {arXiv:2010.12279 [astro-ph.IM]} \BibitemShut {NoStop}%
\bibitem [{\citenamefont {Abarr}\ \emph {et~al.}(2021)\citenamefont {Abarr} \emph {et~al.}}]{PUEO:2020bnn}%
  \BibitemOpen
  \bibfield  {author} {\bibinfo {author} {\bibfnamefont {Q.}~\bibnamefont {Abarr}} \emph {et~al.} (\bibinfo {collaboration} {PUEO Collaboration}),\ }\bibfield  {title} {\bibinfo {title} {{The Payload for Ultrahigh Energy Observations (PUEO): a white paper}},\ }\href {https://doi.org/10.1088/1748-0221/16/08/P08035} {\bibfield  {journal} {\bibinfo  {journal} {JINST}\ }\textbf {\bibinfo {volume} {16}}\bibfield  {number} {\bibinfo  {number} { (08)},\ \bibinfo {pages} {P08035}},\ }\Eprint {https://arxiv.org/abs/2010.02892} {arXiv:2010.02892 [astro-ph.IM]} \BibitemShut {NoStop}%
\bibitem [{\citenamefont {Southall}\ \emph {et~al.}(2021)\citenamefont {Southall} \emph {et~al.}}]{BEACON:2021fpe}%
  \BibitemOpen
  \bibfield  {author} {\bibinfo {author} {\bibfnamefont {D.}~\bibnamefont {Southall}} \emph {et~al.} (\bibinfo {collaboration} {BEACON Collaboration}),\ }\bibfield  {title} {\bibinfo {title} {{Searching for RF-only triggered cosmic ray events with the high-elevation BEACON prototype}},\ }\href {https://doi.org/10.22323/1.395.1084} {\bibfield  {journal} {\bibinfo  {journal} {PoS}\ }\textbf {\bibinfo {volume} {ICRC2021}},\ \bibinfo {pages} {1084} (\bibinfo {year} {2021})}\BibitemShut {NoStop}%
\bibitem [{\citenamefont {Jin}(2024)}]{jin_2024_13235037}%
  \BibitemOpen
  \bibfield  {author} {\bibinfo {author} {\bibfnamefont {M.}~\bibnamefont {Jin}},\ }\href {https://doi.org/10.5281/zenodo.13235037} {\bibinfo {title} {Charm meson induced double cascades in neutrino telescopes}} (\bibinfo {year} {2024})\BibitemShut {NoStop}%
\bibitem [{\citenamefont {Abbasi}\ \emph {et~al.}(2024{\natexlab{b}})\citenamefont {Abbasi} \emph {et~al.}}]{IceCube:2024fxo}%
  \BibitemOpen
  \bibfield  {author} {\bibinfo {author} {\bibfnamefont {R.}~\bibnamefont {Abbasi}} \emph {et~al.} (\bibinfo {collaboration} {IceCube Collaboration}),\ }\bibfield  {title} {\bibinfo {title} {{Characterization of the astrophysical diffuse neutrino flux using starting track events in IceCube}},\ }\href {https://doi.org/10.1103/PhysRevD.110.022001} {\bibfield  {journal} {\bibinfo  {journal} {Phys. Rev. D}\ }\textbf {\bibinfo {volume} {110}},\ \bibinfo {pages} {022001} (\bibinfo {year} {2024}{\natexlab{b}})},\ \Eprint {https://arxiv.org/abs/2402.18026} {arXiv:2402.18026 [astro-ph.HE]} \BibitemShut {NoStop}%
\bibitem [{\citenamefont {Fedynitch}\ \emph {et~al.}(2015)\citenamefont {Fedynitch}, \citenamefont {Engel}, \citenamefont {Gaisser}, \citenamefont {Riehn},\ and\ \citenamefont {Stanev}}]{Fedynitch:2015zma}%
  \BibitemOpen
  \bibfield  {author} {\bibinfo {author} {\bibfnamefont {A.}~\bibnamefont {Fedynitch}}, \bibinfo {author} {\bibfnamefont {R.}~\bibnamefont {Engel}}, \bibinfo {author} {\bibfnamefont {T.~K.}\ \bibnamefont {Gaisser}}, \bibinfo {author} {\bibfnamefont {F.}~\bibnamefont {Riehn}},\ and\ \bibinfo {author} {\bibfnamefont {T.}~\bibnamefont {Stanev}},\ }\bibfield  {title} {\bibinfo {title} {{Calculation of conventional and prompt lepton fluxes at very high energy}},\ }\href {https://doi.org/10.1051/epjconf/20159908001} {\bibfield  {journal} {\bibinfo  {journal} {EPJ Web Conf.}\ }\textbf {\bibinfo {volume} {99}},\ \bibinfo {pages} {08001} (\bibinfo {year} {2015})},\ \Eprint {https://arxiv.org/abs/1503.00544} {arXiv:1503.00544 [hep-ph]} \BibitemShut {NoStop}%
\bibitem [{\citenamefont {Riehn}\ \emph {et~al.}(2018)\citenamefont {Riehn}, \citenamefont {Dembinski}, \citenamefont {Engel}, \citenamefont {Fedynitch}, \citenamefont {Gaisser},\ and\ \citenamefont {Stanev}}]{Riehn:2017mfm}%
  \BibitemOpen
  \bibfield  {author} {\bibinfo {author} {\bibfnamefont {F.}~\bibnamefont {Riehn}}, \bibinfo {author} {\bibfnamefont {H.~P.}\ \bibnamefont {Dembinski}}, \bibinfo {author} {\bibfnamefont {R.}~\bibnamefont {Engel}}, \bibinfo {author} {\bibfnamefont {A.}~\bibnamefont {Fedynitch}}, \bibinfo {author} {\bibfnamefont {T.~K.}\ \bibnamefont {Gaisser}},\ and\ \bibinfo {author} {\bibfnamefont {T.}~\bibnamefont {Stanev}},\ }\bibfield  {title} {\bibinfo {title} {{The hadronic interaction model SIBYLL 2.3c and Feynman scaling}},\ }\href {https://doi.org/10.22323/1.301.0301} {\bibfield  {journal} {\bibinfo  {journal} {PoS}\ }\textbf {\bibinfo {volume} {ICRC2017}},\ \bibinfo {pages} {301} (\bibinfo {year} {2018})},\ \Eprint {https://arxiv.org/abs/1709.07227} {arXiv:1709.07227 [hep-ph]} \BibitemShut {NoStop}%
\bibitem [{\citenamefont {Gaisser}\ \emph {et~al.}(2013)\citenamefont {Gaisser}, \citenamefont {Stanev},\ and\ \citenamefont {Tilav}}]{Gaisser:2013bla}%
  \BibitemOpen
  \bibfield  {author} {\bibinfo {author} {\bibfnamefont {T.~K.}\ \bibnamefont {Gaisser}}, \bibinfo {author} {\bibfnamefont {T.}~\bibnamefont {Stanev}},\ and\ \bibinfo {author} {\bibfnamefont {S.}~\bibnamefont {Tilav}},\ }\bibfield  {title} {\bibinfo {title} {{Cosmic ray rnergy spectrum from measurements of air showers}},\ }\href {https://doi.org/10.1007/s11467-013-0319-7} {\bibfield  {journal} {\bibinfo  {journal} {Front. Phys. (Beijing)}\ }\textbf {\bibinfo {volume} {8}},\ \bibinfo {pages} {748} (\bibinfo {year} {2013})},\ \Eprint {https://arxiv.org/abs/1303.3565} {arXiv:1303.3565 [astro-ph.HE]} \BibitemShut {NoStop}%
\bibitem [{\citenamefont {Aartsen}\ \emph {et~al.}(2019)\citenamefont {Aartsen} \emph {et~al.}}]{IceCube:2018pgc}%
  \BibitemOpen
  \bibfield  {author} {\bibinfo {author} {\bibfnamefont {M.~G.}\ \bibnamefont {Aartsen}} \emph {et~al.} (\bibinfo {collaboration} {IceCube Collaboration}),\ }\bibfield  {title} {\bibinfo {title} {{Measurements using the inelasticity distribution of multi-TeV neutrino interactions in IceCube}},\ }\href {https://doi.org/10.1103/PhysRevD.99.032004} {\bibfield  {journal} {\bibinfo  {journal} {Phys. Rev. D}\ }\textbf {\bibinfo {volume} {99}},\ \bibinfo {pages} {032004} (\bibinfo {year} {2019})},\ \Eprint {https://arxiv.org/abs/1808.07629} {arXiv:1808.07629 [hep-ex]} \BibitemShut {NoStop}%
\bibitem [{\citenamefont {Schneider}\ \emph {et~al.}(2025)\citenamefont {Schneider}, \citenamefont {Kamp},\ and\ \citenamefont {Wen}}]{Schneider:2024eej}%
  \BibitemOpen
  \bibfield  {author} {\bibinfo {author} {\bibfnamefont {A.}~\bibnamefont {Schneider}}, \bibinfo {author} {\bibfnamefont {N.~W.}\ \bibnamefont {Kamp}},\ and\ \bibinfo {author} {\bibfnamefont {A.~Y.}\ \bibnamefont {Wen}},\ }\bibfield  {title} {\bibinfo {title} {{SIREN: an open-source neutrino injection toolkit}},\ }\href {https://doi.org/10.1016/j.cpc.2025.109799} {\bibfield  {journal} {\bibinfo  {journal} {Comput. Phys. Commun.}\ }\textbf {\bibinfo {volume} {316}},\ \bibinfo {pages} {109799} (\bibinfo {year} {2025})},\ \Eprint {https://arxiv.org/abs/2406.01745} {arXiv:2406.01745 [hep-ex]} \BibitemShut {NoStop}%
\bibitem [{\citenamefont {Schonert}\ \emph {et~al.}(2009)\citenamefont {Schonert}, \citenamefont {Gaisser}, \citenamefont {Resconi},\ and\ \citenamefont {Schulz}}]{Schonert:2008is}%
  \BibitemOpen
  \bibfield  {author} {\bibinfo {author} {\bibfnamefont {S.}~\bibnamefont {Schonert}}, \bibinfo {author} {\bibfnamefont {T.~K.}\ \bibnamefont {Gaisser}}, \bibinfo {author} {\bibfnamefont {E.}~\bibnamefont {Resconi}},\ and\ \bibinfo {author} {\bibfnamefont {O.}~\bibnamefont {Schulz}},\ }\bibfield  {title} {\bibinfo {title} {{Vetoing atmospheric neutrinos in a high energy neutrino telescope}},\ }\href {https://doi.org/10.1103/PhysRevD.79.043009} {\bibfield  {journal} {\bibinfo  {journal} {Phys. Rev. D}\ }\textbf {\bibinfo {volume} {79}},\ \bibinfo {pages} {043009} (\bibinfo {year} {2009})},\ \Eprint {https://arxiv.org/abs/0812.4308} {arXiv:0812.4308 [astro-ph]} \BibitemShut {NoStop}%
\bibitem [{\citenamefont {Gaisser}\ \emph {et~al.}(2014)\citenamefont {Gaisser}, \citenamefont {Jero}, \citenamefont {Karle},\ and\ \citenamefont {van Santen}}]{Gaisser:2014bja}%
  \BibitemOpen
  \bibfield  {author} {\bibinfo {author} {\bibfnamefont {T.~K.}\ \bibnamefont {Gaisser}}, \bibinfo {author} {\bibfnamefont {K.}~\bibnamefont {Jero}}, \bibinfo {author} {\bibfnamefont {A.}~\bibnamefont {Karle}},\ and\ \bibinfo {author} {\bibfnamefont {J.}~\bibnamefont {van Santen}},\ }\bibfield  {title} {\bibinfo {title} {{Generalized self-veto probability for atmospheric neutrinos}},\ }\href {https://doi.org/10.1103/PhysRevD.90.023009} {\bibfield  {journal} {\bibinfo  {journal} {Phys. Rev. D}\ }\textbf {\bibinfo {volume} {90}},\ \bibinfo {pages} {023009} (\bibinfo {year} {2014})},\ \Eprint {https://arxiv.org/abs/1405.0525} {arXiv:1405.0525 [astro-ph.HE]} \BibitemShut {NoStop}%
\bibitem [{\citenamefont {Arg{\"u}elles}\ \emph {et~al.}(2018)\citenamefont {Arg{\"u}elles}, \citenamefont {Palomares-Ruiz}, \citenamefont {Schneider}, \citenamefont {Wille},\ and\ \citenamefont {Yuan}}]{Arguelles:2018awr}%
  \BibitemOpen
  \bibfield  {author} {\bibinfo {author} {\bibfnamefont {C.~A.}\ \bibnamefont {Arg{\"u}elles}}, \bibinfo {author} {\bibfnamefont {S.}~\bibnamefont {Palomares-Ruiz}}, \bibinfo {author} {\bibfnamefont {A.}~\bibnamefont {Schneider}}, \bibinfo {author} {\bibfnamefont {L.}~\bibnamefont {Wille}},\ and\ \bibinfo {author} {\bibfnamefont {T.}~\bibnamefont {Yuan}},\ }\bibfield  {title} {\bibinfo {title} {{Unified atmospheric neutrino passing fractions for large-scale neutrino telescopes}},\ }\href {https://doi.org/10.1088/1475-7516/2018/07/047} {\bibfield  {journal} {\bibinfo  {journal} {JCAP}\ }\textbf {\bibinfo {volume} {07}},\ \bibinfo {pages} {047}},\ \Eprint {https://arxiv.org/abs/1805.11003} {arXiv:1805.11003 [hep-ph]} \BibitemShut {NoStop}%
\bibitem [{\citenamefont {Weigel}(2025)}]{weigel_2025_15690410}%
  \BibitemOpen
  \bibfield  {author} {\bibinfo {author} {\bibfnamefont {P.}~\bibnamefont {Weigel}} (\bibinfo {collaboration} {IceCube Collaboration}),\ }\href {https://doi.org/10.5281/zenodo.15690410} {\bibinfo {title} {Machine learning techniques for neutrino reconstructions in icecube}} (\bibinfo {year} {2025})\BibitemShut {NoStop}%
\bibitem [{\citenamefont {Balagopal~V.}\ \emph {et~al.}(2025)\citenamefont {Balagopal~V.} \emph {et~al.}}]{IceCube:2025uyt}%
  \BibitemOpen
  \bibfield  {author} {\bibinfo {author} {\bibfnamefont {A.}~\bibnamefont {Balagopal~V.}} \emph {et~al.} (\bibinfo {collaboration} {IceCube Collaboration}),\ }\bibfield  {title} {\bibinfo {title} {{Measurement of the three-flavor composition of astrophysical neutrinos with contained IceCube events}},\ }\href {https://doi.org/10.22323/1.501.0983} {\bibfield  {journal} {\bibinfo  {journal} {PoS}\ }\textbf {\bibinfo {volume} {ICRC2025}},\ \bibinfo {pages} {983} (\bibinfo {year} {2025})},\ \Eprint {https://arxiv.org/abs/2507.07212} {arXiv:2507.07212 [astro-ph.HE]} \BibitemShut {NoStop}%
\bibitem [{\citenamefont {Abbasi}\ \emph {et~al.}(2021)\citenamefont {Abbasi} \emph {et~al.}}]{IceCube:2020wum}%
  \BibitemOpen
  \bibfield  {author} {\bibinfo {author} {\bibfnamefont {R.}~\bibnamefont {Abbasi}} \emph {et~al.} (\bibinfo {collaboration} {IceCube Collaboration}),\ }\bibfield  {title} {\bibinfo {title} {{The IceCube high-energy starting event sample: Description and flux characterization with 7.5 years of data}},\ }\href {https://doi.org/10.1103/PhysRevD.104.022002} {\bibfield  {journal} {\bibinfo  {journal} {Phys. Rev. D}\ }\textbf {\bibinfo {volume} {104}},\ \bibinfo {pages} {022002} (\bibinfo {year} {2021})},\ \Eprint {https://arxiv.org/abs/2011.03545} {arXiv:2011.03545 [astro-ph.HE]} \BibitemShut {NoStop}%
\bibitem [{\citenamefont {Aartsen}\ \emph {et~al.}(2020)\citenamefont {Aartsen} \emph {et~al.}}]{IceCube:2020acn}%
  \BibitemOpen
  \bibfield  {author} {\bibinfo {author} {\bibfnamefont {M.~G.}\ \bibnamefont {Aartsen}} \emph {et~al.} (\bibinfo {collaboration} {IceCube Collaboration}),\ }\bibfield  {title} {\bibinfo {title} {{Characteristics of the diffuse astrophysical electron and tau neutrino flux with six years of IceCube high energy cascade data}},\ }\href {https://doi.org/10.1103/PhysRevLett.125.121104} {\bibfield  {journal} {\bibinfo  {journal} {Phys. Rev. Lett.}\ }\textbf {\bibinfo {volume} {125}},\ \bibinfo {pages} {121104} (\bibinfo {year} {2020})},\ \Eprint {https://arxiv.org/abs/2001.09520} {arXiv:2001.09520 [astro-ph.HE]} \BibitemShut {NoStop}%
\bibitem [{\citenamefont {Abbasi}\ \emph {et~al.}(2022{\natexlab{b}})\citenamefont {Abbasi} \emph {et~al.}}]{Abbasi:2021qfz}%
  \BibitemOpen
  \bibfield  {author} {\bibinfo {author} {\bibfnamefont {R.}~\bibnamefont {Abbasi}} \emph {et~al.} (\bibinfo {collaboration} {IceCube Collaboration}),\ }\bibfield  {title} {\bibinfo {title} {{Improved characterization of the astrophysical muon{\textendash}neutrino flux with 9.5 years of IceCube data}},\ }\href {https://doi.org/10.3847/1538-4357/ac4d29} {\bibfield  {journal} {\bibinfo  {journal} {Astrophys. J.}\ }\textbf {\bibinfo {volume} {928}},\ \bibinfo {pages} {50} (\bibinfo {year} {2022}{\natexlab{b}})},\ \Eprint {https://arxiv.org/abs/2111.10299} {arXiv:2111.10299 [astro-ph.HE]} \BibitemShut {NoStop}%
\bibitem [{\citenamefont {Plestid}\ and\ \citenamefont {Zhou}(2025)}]{Plestid:2024bva}%
  \BibitemOpen
  \bibfield  {author} {\bibinfo {author} {\bibfnamefont {R.}~\bibnamefont {Plestid}}\ and\ \bibinfo {author} {\bibfnamefont {B.}~\bibnamefont {Zhou}},\ }\bibfield  {title} {\bibinfo {title} {{Final state radiation from high and ultrahigh energy neutrino interactions}},\ }\href {https://doi.org/10.1103/PhysRevD.111.043007} {\bibfield  {journal} {\bibinfo  {journal} {Phys. Rev. D}\ }\textbf {\bibinfo {volume} {111}},\ \bibinfo {pages} {043007} (\bibinfo {year} {2025})},\ \Eprint {https://arxiv.org/abs/2403.07984} {arXiv:2403.07984 [hep-ph]} \BibitemShut {NoStop}%
\bibitem [{\citenamefont {Aartsen}\ \emph {et~al.}(2021)\citenamefont {Aartsen} \emph {et~al.}}]{IceCube-Gen2:2020qha}%
  \BibitemOpen
  \bibfield  {author} {\bibinfo {author} {\bibfnamefont {M.~G.}\ \bibnamefont {Aartsen}} \emph {et~al.} (\bibinfo {collaboration} {IceCube-Gen2 Collaboration}),\ }\bibfield  {title} {\bibinfo {title} {{IceCube-Gen2: the window to the extreme Universe}},\ }\href {https://doi.org/10.1088/1361-6471/abbd48} {\bibfield  {journal} {\bibinfo  {journal} {J. Phys. G}\ }\textbf {\bibinfo {volume} {48}},\ \bibinfo {pages} {060501} (\bibinfo {year} {2021})},\ \Eprint {https://arxiv.org/abs/2008.04323} {arXiv:2008.04323 [astro-ph.HE]} \BibitemShut {NoStop}%
\bibitem [{\citenamefont {Abbasi}\ \emph {et~al.}(2022{\natexlab{c}})\citenamefont {Abbasi} \emph {et~al.}}]{IceCube:2022der}%
  \BibitemOpen
  \bibfield  {author} {\bibinfo {author} {\bibfnamefont {R.}~\bibnamefont {Abbasi}} \emph {et~al.} (\bibinfo {collaboration} {IceCube Collaboration}),\ }\bibfield  {title} {\bibinfo {title} {{Evidence for neutrino emission from the nearby active galaxy NGC 1068}},\ }\href {https://doi.org/10.1126/science.abg3395} {\bibfield  {journal} {\bibinfo  {journal} {Science}\ }\textbf {\bibinfo {volume} {378}},\ \bibinfo {pages} {538} (\bibinfo {year} {2022}{\natexlab{c}})},\ \Eprint {https://arxiv.org/abs/2211.09972} {arXiv:2211.09972 [astro-ph.HE]} \BibitemShut {NoStop}%
\bibitem [{\citenamefont {Thompson}\ \emph {et~al.}(2025)\citenamefont {Thompson} \emph {et~al.}}]{Thompson:2025tng}%
  \BibitemOpen
  \bibfield  {author} {\bibinfo {author} {\bibfnamefont {W.}~\bibnamefont {Thompson}} \emph {et~al.} (\bibinfo {collaboration} {TAMBO Collaboration}),\ }\bibfield  {title} {\bibinfo {title} {{Current status of TAMBO: realizing PeV neutrino astronomy with a cost-effective observatory}},\ }\href {https://doi.org/10.22323/1.501.1194} {\bibfield  {journal} {\bibinfo  {journal} {PoS}\ }\textbf {\bibinfo {volume} {ICRC2025}},\ \bibinfo {pages} {1194} (\bibinfo {year} {2025})}\BibitemShut {NoStop}%
\bibitem [{\citenamefont {Cooper-Sarkar}\ \emph {et~al.}(2011)\citenamefont {Cooper-Sarkar}, \citenamefont {Mertsch},\ and\ \citenamefont {Sarkar}}]{Cooper-Sarkar:2011jtt}%
  \BibitemOpen
  \bibfield  {author} {\bibinfo {author} {\bibfnamefont {A.}~\bibnamefont {Cooper-Sarkar}}, \bibinfo {author} {\bibfnamefont {P.}~\bibnamefont {Mertsch}},\ and\ \bibinfo {author} {\bibfnamefont {S.}~\bibnamefont {Sarkar}},\ }\bibfield  {title} {\bibinfo {title} {{The high energy neutrino cross-section in the Standard Model and its uncertainty}},\ }\href {https://doi.org/10.1007/JHEP08(2011)042} {\bibfield  {journal} {\bibinfo  {journal} {JHEP}\ }\textbf {\bibinfo {volume} {08}},\ \bibinfo {pages} {042}},\ \Eprint {https://arxiv.org/abs/1106.3723} {arXiv:1106.3723 [hep-ph]} \BibitemShut {NoStop}%
\bibitem [{\citenamefont {Niess}(2023)}]{Niess:2022msc}%
  \BibitemOpen
  \bibfield  {author} {\bibinfo {author} {\bibfnamefont {V.}~\bibnamefont {Niess}},\ }\bibfield  {title} {\bibinfo {title} {{Alouette: Yet another encapsulated TAUOLA, but revertible}},\ }\href {https://doi.org/10.1016/j.cpc.2022.108508} {\bibfield  {journal} {\bibinfo  {journal} {Comput. Phys. Commun.}\ }\textbf {\bibinfo {volume} {282}},\ \bibinfo {pages} {108508} (\bibinfo {year} {2023})},\ \Eprint {https://arxiv.org/abs/2208.11914} {arXiv:2208.11914 [physics.comp-ph]} \BibitemShut {NoStop}%
\bibitem [{\citenamefont {Jadach}\ \emph {et~al.}(1990)\citenamefont {Jadach}, \citenamefont {Kuhn},\ and\ \citenamefont {Was}}]{Jadach:1990mz}%
  \BibitemOpen
  \bibfield  {author} {\bibinfo {author} {\bibfnamefont {S.}~\bibnamefont {Jadach}}, \bibinfo {author} {\bibfnamefont {J.~H.}\ \bibnamefont {Kuhn}},\ and\ \bibinfo {author} {\bibfnamefont {Z.}~\bibnamefont {Was}},\ }\bibfield  {title} {\bibinfo {title} {{TAUOLA: a library of Monte Carlo programs to simulate decays of polarized tau leptons}},\ }\href {https://doi.org/10.1016/0010-4655(91)90038-M} {\bibfield  {journal} {\bibinfo  {journal} {Comput. Phys. Commun.}\ }\textbf {\bibinfo {volume} {64}},\ \bibinfo {pages} {275} (\bibinfo {year} {1990})}\BibitemShut {NoStop}%
\bibitem [{\citenamefont {Jadach}\ \emph {et~al.}(1993)\citenamefont {Jadach}, \citenamefont {Was}, \citenamefont {Decker},\ and\ \citenamefont {Kuhn}}]{Jadach:1993hs}%
  \BibitemOpen
  \bibfield  {author} {\bibinfo {author} {\bibfnamefont {S.}~\bibnamefont {Jadach}}, \bibinfo {author} {\bibfnamefont {Z.}~\bibnamefont {Was}}, \bibinfo {author} {\bibfnamefont {R.}~\bibnamefont {Decker}},\ and\ \bibinfo {author} {\bibfnamefont {J.~H.}\ \bibnamefont {Kuhn}},\ }\bibfield  {title} {\bibinfo {title} {{The tau decay library TAUOLA: Version 2.4}},\ }\href {https://doi.org/10.1016/0010-4655(93)90061-G} {\bibfield  {journal} {\bibinfo  {journal} {Comput. Phys. Commun.}\ }\textbf {\bibinfo {volume} {76}},\ \bibinfo {pages} {361} (\bibinfo {year} {1993})}\BibitemShut {NoStop}%
\bibitem [{\citenamefont {IceCube}(2023)}]{NuFluxRepository}%
  \BibitemOpen
  \bibfield  {author} {\bibinfo {author} {\bibnamefont {IceCube}},\ }\href@noop {} {\bibinfo {title} {\texttt{Nuflux code}}},\ \bibinfo {howpublished} {\url{https://github.com/icecube/nuflux}} (\bibinfo {year} {2023})\BibitemShut {NoStop}%
\end{thebibliography}%
\clearpage


\newpage

\onecolumngrid
\appendix

\ifx \standalonesupplemental\undefined
\setcounter{page}{1}
\setcounter{figure}{0}
\setcounter{table}{0}
\setcounter{equation}{0}
\fi

\renewcommand{\thepage}{Supplemental Methods and Tables -- S\arabic{page}}
\renewcommand{\figurename}{SUPPL. FIG.}
\renewcommand{\tablename}{SUPPL. TABLE}

\renewcommand{\theequation}{A\arabic{equation}}

\section{\Large{Supplemental Material}}

\subsection{Details on the analysis method}

In the following, we provide additional details of the analysis beyond those described in the main text.

To simulate a population of astrophysical starting tracks in IceCube, we consider a cubic kilometer of ice, approximated in \texttt{SIREN}~\cite{Schneider:2024eej} as a cylinder. Within this volume, we generate (anti)neutrino–nucleon interactions using the CSMS cross sections for both muon and tau neutrinos~\cite{Cooper-Sarkar:2011jtt}, assuming an $E_\nu^{-2.5}$ energy spectrum, which can be later reweighted to any arbitrary spectral index. 
The true inelasticity, $y$, can be computed from the true energies of the injected neutrino and the outgoing lepton.

For tau neutrino interactions, the tau is decayed to the muonic channel using the \texttt{alouette} wrapper~\cite{Niess:2022msc} of the \texttt{tauola} library~\cite{Jadach:1990mz,Jadach:1993hs}, assuming the true tau energy from the \texttt{SIREN} interaction and the full tau polarization. 
The corresponding event weights are scaled by a factor of 0.17 to account for the muonic branching ratio. 
For both muon and tau channels, the true visible inelasticity, $y_{\textrm{vis}}$, is calculated using the muon energy and the hadronic cascade energy. 
At these energies, the tau propagation length is negligible, so the track energy is well approximated by the muon energy. In this manner, $y_\mathrm{vis} = y$ for tracks produced by muon neutrinos and $y_\mathrm{vis} > y$ for tracks produced by tau neutrinos.

At this stage, the simulated events have correct relative weights but are not absolutely normalized. 
The normalization is obtained by integrating the product of the flux, energy range, effective area, detector livetime, zenith range, and solid angle. 
The effective area is that of Ref.~\cite{IceCube:2024fxo}, which includes detector selection and reconstruction efficiencies.
This yields the expected total number of events, allowing the simulated sample to be properly normalized.

The energy spectrum of the atmospheric neutrino background is computed with the \texttt{MCEq} toolkit~\cite{Fedynitch:2015zma} using the \texttt{SIBYLL2.3c} hadronic interaction model~\cite{Riehn:2017mfm} and the \texttt{HillasGaisser2012} cosmic ray flux~\cite{Gaisser:2013bla}.
These flux results can be read easily via the package \texttt{NuFlux}~\cite{NuFluxRepository}.
A similar integral is performed to compute the correct normalization. 
For the atmospheric neutrino background calculation, we also subject the flux to the self-veto correction~\cite{Schonert:2008is, Gaisser:2014bja, Arguelles:2018awr}, which accounts for the fact that some of these neutrino events will be vetoed by a muon coming from the same shower. 
This is most relevant for more vertical, downgoing, events and events at the highest energies. 
We model the self-veto correction as an energy- and zenith-dependent attenuation of the flux using the tables in Ref.~\cite{Arguelles:2018awr}.

For the diffuse flux case, the weighted events of each flux and flavor are then binned in energy, zenith angle and $y_\mathrm{vis}$, and a binned likelihood analysis is performed.
The events are split into two energy bins (10--100~TeV and 100--1000~TeV); three zenith bins, $-1 < \cos\theta_z < 0$ (upgoing), $0 < \cos\theta_z < 0.5$ (horizontal), and $0.5 < \cos\theta_z < 1$ (downgoing); and five $y_\mathrm{vis}$ bins equally spaced in $[0,1]$.
We use a Poisson likelihood over all bins, $\mathcal{L} = \prod_i \text{Poisson}(d_i | n_i)$, where $d_i$ is the number of mock data events in bin $i$ and $n_i$ is the expected number of counts in than bin. Both include the atmospheric neutrino background and the astrophysical contribution, which depends on $R_{\tau\mu}$, on the astrophysical normalization $N_\phi$ and on the astrophysical spectral index $\gamma$. 
On another hand, since the contribution from atmospheric muons is expected to be subdominant~\cite{IceCube:2018pgc} and concentrated in the lowest-$y_\mathrm{vis}$ region, its impact on the analysis is expected to be negligible and thus, we do not include it.
The benchmark (null) astrophysical scenario assumes a flavor ratio at the detector $(\nu_e:\nu_\mu:\nu_\tau) = (1:1:1)$, i.e., $R_{\tau\mu} = 1$, with equal contributions from neutrinos and antineutrinos. Since the cross sections of different flavors are almost the same at the energies we consider, the number of expected astrophysical events in a given bin approximately depends on $R_{\tau\mu}$, on the overall normalization and on the spectral index through the combination $N_\phi \, \left(1 + 0.17 \, R_{\tau\mu}\right) \, \phi_{\nu_\mu}(\gamma)$ for neutrinos, and similarly for antineutrinos, $N_\phi \, \left(1 + 0.17 \, R_{\tau\mu}\right) \, \phi_{\bar\nu_\mu}(\gamma)$, where the factor of $0.17$ accounts for the tau-to-muon decay branching ratio, and $\phi_{\nu_\mu} = \phi_{\bar\nu_\mu} = \frac{1}{6} \, \phi_\mathrm{astro}(E_\nu) \, \left(E_\nu/100 \, \mathrm{TeV}\right)^{-\gamma+2.58}$, with $\phi_\mathrm{astro}(E_\nu)$ given in Eq.~(\ref{eq:astroflux})~\cite{IceCube:2024fxo}. 

In the fit, we include a Gaussian prior on $N_\phi$ of 15\%, centered at 1, and no prior on $\gamma$, and then profile over $N_\phi$ and $\gamma$ to obtain the 68\% CL region in $R_{\tau\mu}$. 
After profiling over $N_\phi$ and $\gamma$, we obtain the log-likelihood ratio,
\begin{equation}
   - 2 \ln \lambda(R_{\tau\mu}) = \min_{N_\phi, \gamma} \, \left[ 2 \sum_i \left(n_i(R_{\tau\mu}; N_\phi, \gamma) - d_i + d_i \log\frac{d_i}{n_i(R_{\tau\mu}; N_\phi, \gamma)} \right) + \frac{(N_\phi -1)^2}{\sigma_{N_\phi}^2} \right] ~,
\end{equation}
where $n_i(R_{\tau\mu}; N_\phi, \gamma) = n_{\mathrm{astro},i}(R_{\tau\mu}; N_\phi, \gamma) + n_{\mathrm{atmos},i}$, $d_i = n_{\mathrm{astro},i}(R_{\tau\mu}=1; N_\phi=1, \gamma=2.58) + n_{\mathrm{atmos},i}$, and $\sigma_{N_\phi} = 0.15$. The resulting sensitivity on $R_{\tau\mu}$ is obtained using Wilks' theorem, assuming that the log-likehood ratio $- 2 \ln\lambda(R_{\tau\mu})$ follows a $\chi^2$ distribution with one degree of freedom.

The results of this analysis are shown in Fig.~\ref{fig:sensitivity_diffuse}. Additionally, Suppl. Fig.~\ref{fig:sensitivity_diffuse_wtheory} shows the effect of dropping the prior assumption on $N_\phi$. As shown in the figure, the uncertainty on $R_{\tau\mu}$ increases slightly, although the effect is not very significant. In Fig.~\ref{fig:y-resolutions}, we repeat the analysis in Fig.~\ref{fig:sensitivity_diffuse} for different values of $\sigma_\mathrm{sys}$, but using ten bins in $y_\mathrm{vis}$ instead of five.

For the point-source scenario shown in Fig.~\ref{fig:sensitivity_PS}, we do not vary $N_\phi$ and $\gamma$. We simply fix $\gamma = 2.58$ for all cases and fix $N_\phi$ at various values corresponding to different $N_\mathrm{astro}$. 
The sensitivity is then evaluated with the same likelihood assuming just one zenith angle bin and adding an energy bin, (1--10)~TeV. 
We have checked that, for all directions in the sky, the expected number of atmospheric neutrino events in 10 years is below $1/\textrm{deg}^2$. 
Therefore, this background can be considered negligible for the purposes of this analysis. As a consequence, although we show results for $\cos\theta_z = 0.5$, the source location does not have a significant impact on the final sensitivity.

\subsection{Allowed flavor ratio from different sources} 

\begin{figure}[t!]%
    \centering
    \includegraphics[width=0.7\linewidth]{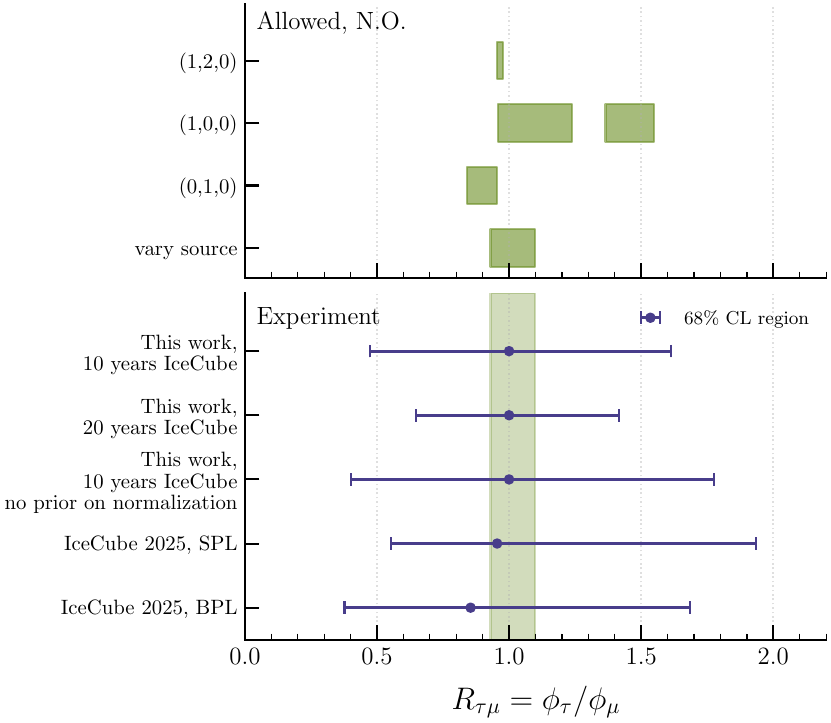}
    \caption{\textit{Projected all-sky flavor measurement sensitivity, with additional allowed regions.} 
    We show the 68\%~CL allowed ranges of $R_{\tau\mu}$ for our simulated all-sky sample of astrophysical neutrinos separated into zenith, inelasticity, and energy bins, the same result of Fig.~\ref{fig:sensitivity_diffuse}.
    In addition, we also add the sensitivity without imposing a prior on the total normalization $N_\phi$.
    The vertical green bands indicate the 68\% highest-density interval of $R_{\tau\mu}$ given the NuFit 2024 mixing parameter allowed ranges at 68\%~CL~\cite{Esteban:2024eli}, but also broken up into different source production hypotheses. 
    The ``vary source'' one corresponds to the scan over source production $(f_e,1-f_e,0)$ with $f_e \in [0,1]$, and is the allowed region that the experimental limits are compared to. 
    Details on the binning are described in the text and can be seen in Fig.~\ref{fig:event_distributions}.
    Like in Fig.~\ref{fig:sensitivity_diffuse}, we also show the profiled flavor measurement result from Ref.~\cite{IceCube:2025uyt}, which includes a fit to both a SPL and a BPL astrophysical flux hypothesis.}%
    \label{fig:sensitivity_diffuse_wtheory}%
\end{figure}

In Suppl. Fig.~\ref{fig:sensitivity_diffuse_wtheory} we provide a more detailed breakdown of the allowed regions under different source production hypotheses, complementing Fig.~\ref{fig:sensitivity_diffuse}. 
In addition to the region obtained by scanning over $(f_e,1-f_e,0)$ — already shown in Fig.~\ref{fig:sensitivity_diffuse} — we include specific scenarios such as $(1,2,0)$ (pion decay), $(0,1,0)$ (muon-damped), and $(1,0,0)$ (neutron decay). 
All regions are derived using the same procedure, just with a different initial source hypothesis. We adopt the NuFit 6.0~\cite{Esteban:2024eli} likelihood landscape as a prior on the mixing parameters, sample from this prior, and map the resulting distributions into $R_{\tau\mu}$ space, in the same way described in the main text. 
The green bands indicate the 68\% highest-density intervals, i.e., the regions containing 68\% of the probability mass around the mode.

\newpage

\subsection{Dependence on the energy spectrum}

\begin{figure}[t!]%
    \centering
    \includegraphics[width=0.7\linewidth]{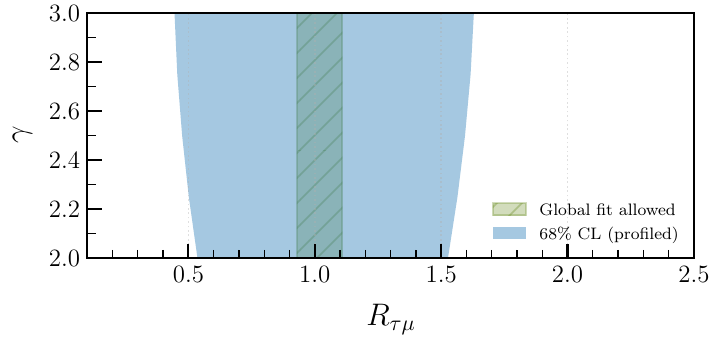}
    \caption{\textit{Dependence of sensitivity on the spectral index, $\gamma$, of the diffuse flux with spectrum $E^{-\gamma}$.}
    We show the allowed region of $R_{\tau\mu}$ as a function of the spectral index of the diffuse flux, with the normalization at the null point of $R_{\tau\mu}=1$ fixed by the best-fit flux value of Ref.~\cite{IceCube:2024fxo}.
    The value assumed for the main result of Fig.~\ref{fig:sensitivity_diffuse} is $\gamma=2.58$. 
    The range of $\gamma$ is designed to span the approximate range of most recent spectral index measurement results.
    }%
    \label{fig:spectral_index_scan}%
\end{figure}

In Suppl. Fig.~\ref{fig:spectral_index_scan} we show the result of the $R_{\tau\mu}$ sensitivity for different values of the assumed astrophysical power-law spectral index, $\gamma$, for the case of the diffuse astrophysical neutrino flux, $E^{-\gamma}$.
To obtain this, we assume a true scenario of a given spectral index and fit to each scenario, done in the same way as in Fig.~\ref{fig:sensitivity_diffuse} by profiling over $N_\phi$ and $\gamma$ to recover an 68\% CL region of $R_{\tau\mu}$.
The overall robustness of the sensitivity, which only varies by less than 0.2 units of $R_{\tau\mu}$ over $\gamma\in[2.25, 2.75]$, underlies the validity of the results in most realistic astrophysical flux scenarios.

\end{document}